\documentclass[prb,showpacs,twocolumn,floatfix]{revtex4}
\usepackage{graphicx}
\begin{document}
\def\deltav{{\mbox{\boldmath{$\delta$}}}}
\input{psfig}
\draft
\title{DIMER STATISTICS ON A BETHE LATTICE}
\author{A. B. Harris and Michael Cohen}
\address{Department of Physics and Astronomy,
University of Pennsylvania, Philadelphia, PA, 19104}

\date{\today}

\begin{abstract}
We discuss the exact solutions of various models of the statistics of
dimer coverings of a Bethe lattice. We reproduce the  well-known
exact result for noninteracting hard-core dimers by both a very
simple geometrical argument and a general algebraic formulation
for lattice statistical problems.  The algebraic formulation
enables us to discuss loop corrections for finite dimensional lattices.
For the Bethe lattice we also obtain the exact solution when either
a) the dimers interact via a short-range interaction or b) the
underlying lattice is anisotropic.  We give the exact solution
for a special limit of dimers on a Bethe lattice in a quenched
random potential in which we consider the maximal covering of
dimers on random clusters at site occupation probability $p$.
Surprisingly the partition function for "maximal coverage" on
the Bethe lattice is identical to that for the statistics of
branched polymers when the activity for a monomer unit
is set equal to $-p$. Finally we give an exact solution for
the number of residual vacancies when hard-core dimers are randomly
deposited on a one dimensional lattice.
\end{abstract}
\maketitle

\section{INTRODUCTION}

The statistics of covering a lattice with monomers or dimers has
a long and continuing history in condensed matter physics.  Recently
there has been a revival of interest in this topic in connection
with a number of seemingly unrelated problems
such as quantum fluctuations in Heisenberg antiferromagnets,\cite{AF}
stability and dynamics in granular systems,\cite{GRAINS}
phase transitions in certain complex fluids,\cite{COMPLEX}
dynamics of catalysis on surfaces,\cite{DLA}
and the biophysics of membranes.\cite{CC1} 
Accordingly, we have been led to revisit this problem with the
goals of a) drastically simplifying the derivation of existing
approximations and b) providing a framework within which the
more modern techniques of statistical mechanics can be applied.

\begin{figure}
\psfig{figure=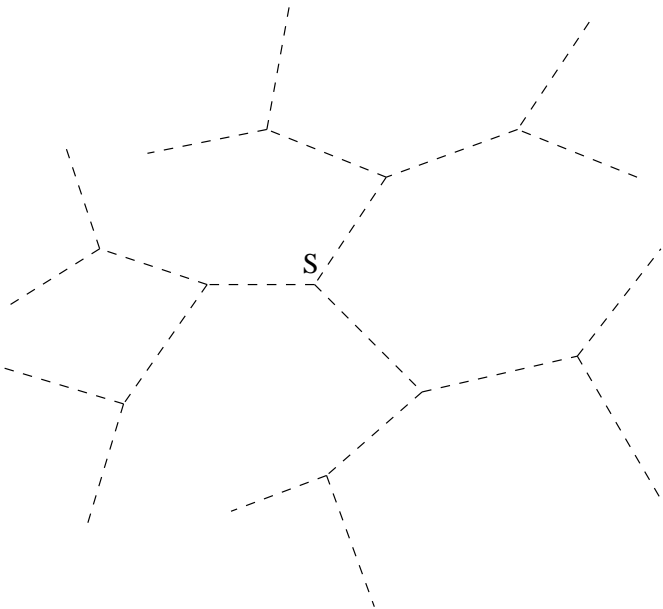}
\caption{A Cayley tree with coordination number $q=3$.
The tree consists of three generations of bonds
starting from the seed site, S.}
\label{CAYFIG}
\end{figure}

The first studies of the statistics of dimer coverings of a lattice
were carried out more that 50 years ago,\cite{B1,B2,B3} obtaining
results analogous to those of the Bethe approximation\cite{HANS,REP,PRW}
for the Ising and Heisenberg models.  At that time, the relation of
this approximation scheme to the structure of the Cayley tree (a
recursive ``lattice,'' an example of which, with coordination number
$q$, is shown in  Fig. \ref{CAYFIG}) was apparently not known.
It was later recognized by Sykes,\cite{MFS} who
apparently first coined the term ``Bethe lattice," that the
Bethe approximation was to be associated with local properties
evaluated near the center of the tree, in order to avoid
surface effects which, for $d$-dimensional hypercubic lattices,
are unphysical.  The pathological effects of the anomalously
large surface were later studied by several authors in the
1970's,\cite{TPE, MAT, ZIT,FALK} but, as was clear from the work of
Fisher and Gaunt in 1964,\cite{MEF} it was the results for local properties
at the center of the infinite tree which could be connected to
those of hypercubic lattices in the limit $d \rightarrow \infty$.
They obtained expansions in powers of $1/d$ for the coefficients of
series expansions in the coupling constant.  However, it was later
shown by the renormalization group\cite{RG} 
that the critical exponents for typical lattice models were
those of mean-field theory for $d>4$.
 
Most of the results of this paper will be obtained for the
Bethe lattice.  We implement the Bethe lattice condition
either by explicitly considering sites far
from the boundary or, alternatively, by using a formulation
appropriate to periodic lattices and then introducing
approximations which become exact when the lattice does not
support any loops.\cite{DOMB}  By initially treating a periodic lattice
we eliminate anomalous surface effects.  We are thus
assured that our results are characteristic of the interior
of the tree and should be associated with what is now
commonly called a Bethe lattice.

Of course, an important aim is to treat real $d$-dimensional lattices.  To this
end there have been a number of papers dealing with series expansions for
the problem of dimer or monomer-dimer coverings of a lattice.  Nagle\cite{JFN}
in 1966 developed a series expansion in powers of the dimer activity $z$ for
a number of two- and three-dimensional lattices.  Longer series were later
obtained obtained by Gaunt\cite{GAUNT}  who exploited the relation between
the dimer problem and the Ising model in a field. Alternative formulations,
based on the generalization of the Mayer cluster expansion\cite{MAYER}
by Rushbrooke and Scoins\cite{SCOINS} have also been given.\cite{DEGREVE}
More recently Brazhnik and Freed\cite{FREED} have given a formulation
suitable not only for dimers but for more complicated entities. Here we
treat the cases of both noninteracting and interacting dimers.
We first address these problems using a simple intuitive geometrical
approach. Then we adopt an algebraic approach based on a transformation
introduced by Shapir\cite{YS} which enables us to develop an expansion for a
$d$-dimensional lattice in which the leading term is the exact result for
the Bethe lattice of the same coordination number $q$. For noninteracting
dimers corrections are obtained in powers of $1/q$ and the activity $z$,
similar to the result of Nagle.\cite{JFN}

Briefly this paper is organized as follows.
In Sec. II we present a simple geometrical derivation of the Bethe
approximation for the statistics of distributing hard-core (but
otherwise noninteracting) dimers on a lattice.  We also give an
alternative algebraic derivation of this result which enables us to
generate loop corrections for the case of a $d$-dimensional periodic
lattice.  We also briefly consider noninteracting dimers on an
anisotropic lattice.  In Sec.  III we consider a model which
includes dimer-dimer interactions.
We derive the Bethe approximation for this model by both the geometrical
and algebraic approaches.  The exact solution of a model in the special
limit of quenched randomness in which each cluster of randomly occupied
sites is maximally covered by noninteracting dimers is given in Sec. IV.
In Sec. V we consider a model of random deposition of hard-core dimers
in one dimension.  We give an exact result for the fraction of sites
which remain vacant after deposition is completed.
Finally, our conclusions are summarized in Sec. VI. In a future
paper we will generalize our approach to treat the statistics of
entities more complicated than dimers.

\section{NONINTERACTING DIMERS ON A BETHE LATTICE}

\subsection{Geometrical Derivation of the Exact Result}

Here we will develop a formula for $\rho$, the average number
of dimers per edge on a lattice in terms of the dimer
chemical potential $\mu$, or preferably in terms of the dimer
"activity" $z \equiv e^{\beta \mu}$, where $\beta=1/(kT)$.
(Interactions between dimers and the underlying lattice as well
as internal degrees of freedom of the dimer are easily included by
a redefinition of $z$.)
We will give what we believe to be the simplest possible ``geometric''
derivation of the well-known\cite{B1,B2,B3} ``Bethe approximation,'' 
which has been shown\cite{CC1} to be surprisingly accurate for
some two dimensional lattices, and which is exact when applied to
the Bethe lattice.  We assume that only dimers can be adsorbed on the lattice,
and that they can be adsorbed only as lying dimers (i.e., an adsorbed
dimer covers two lattice sites).  In the present section we will
assume that there are no interactions between adsorbed dimers except
for the hard-core restriction that two dimers cannot touch the same site.
It is easy to generalize this treatment to
deal with the two-component lattice and also to include the possibility
of adsorption of standing dimers and monomers.\cite{ELSE} 

As stated in the introduction, we initially deal with a periodic
$d$-dimensional lattice and will obtain the results for the Bethe
lattice by introducing an approximation which is exact when the
lattice is tree-like (i. e. it contains no loops).
As a starting point of this discussion we introduce the (unnormalized)
probability that there are $N_D$  dimers on the lattice, $W_N(N_D)z^{N_D}$,
where $W_N(N_D)$ is the number of distinct configurations of $N_D$ dimers
on a lattice of $N$ sites.
The equation of state for this dimer system is the relation between $z$
and $\overline N_D/N$, where $\overline N_D$ is the average value of
$N_D$. In the Bethe approximation\cite{B1,B2,B3}
\begin{eqnarray}
z &=& {2 \over q} {\overline N_D \over N} \left( 1 - {2 \overline N_D \over
qN} \right) \left( 1 - {2\overline N_D \over N} \right)^{-2} \ ,
\label{BETHEEQ} \end{eqnarray}
where $q$ is number of sites which are nearest neighbors of a given site.
To emphasize that this result applies to a Bethe lattice we express it
in terms of densities such as $\rho \equiv 2\overline N_D/(qN)$, the fraction
of bonds (a bond is an edge connecting two sites) covered by a dimer or
$q\rho$, the fraction of sites
which are covered by a dimer. Stated alternatively, $\rho$ is the
probability that a given bond is covered by a dimer and $q \rho$ is
the probability that a given site is covered.  In terms of these variables
the equation of state is
\begin{eqnarray}
z &=& \rho (1-\rho )(1-q\rho )^{-2} 
\label{BETHE1EQ} \end{eqnarray}
or, equivalently,
\begin{eqnarray}
\rho &=& {1 + 2qz - \sqrt{1+4\sigma z} \over 2(1 + q^2 z)} \ ,
\label{BETHE2EQ} \end{eqnarray}
where $\sigma=q-1$. 

\begin{figure}
\psfig{figure=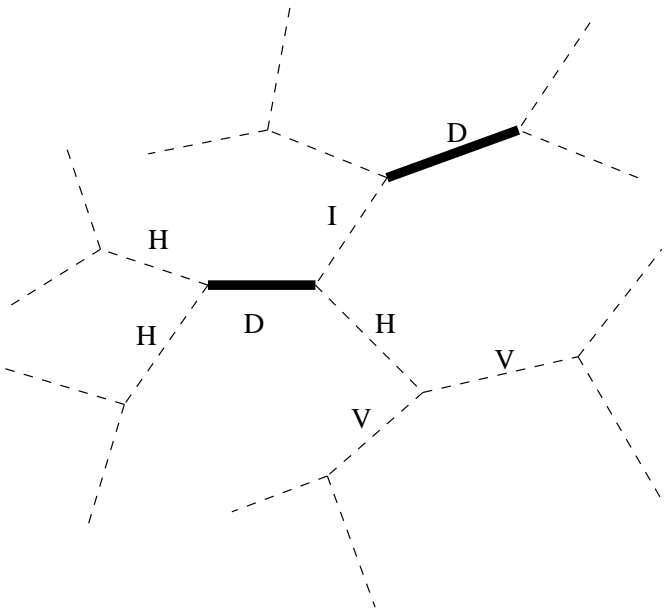}
\caption{Classification of bonds in a lattice partially covered
by dimers.  As indicated in the text D labels a dimer, V a dimer
vacancy, I an interacting bond, and H a half bond.}
\label{FIG1}
\end{figure}

For this discussion it is convenient to define some terminology to
describe configurations of dimers on a lattice.  In Fig. \ref{FIG1}
we show a small section of a lattice where one sees occupied bonds (D)
on which dimers are placed
and unoccupied bonds, which can be of three types depending on
whether or not the two sites of the bond are covered by dimers.
If both sites of an unoccupied bond are uncovered, the bond is called
a ``dimer vacancy,'' (V).  If both sites of the unoccupied bond are covered,
the bond is called (in anticipation of Sec. III where we allow dimer-dimer
interactions) an ``interacting bond'' (I), and if only one site of the
unoccupied bond is covered, the bond is called a ``half bond'' (H).
We first observe that in the thermodynamic limit the average and most
probable values of $N_D$ differ negligibly.  Thus we may say that
the average number of dimers $\overline N_D$ is
the value of $N_D$ for which the probability $W_N(N_D)z^{N_D}$ is maximal.
The condition that this quantity be stationary is
\begin{eqnarray}
{W_N(\overline N_D+1) \over W_N(\overline N_D)} &=& {1 \over z} \ .
\end{eqnarray}
Consider the left-hand side of this equation.  From any configuration of $N_D$
dimers we can obtain a configuration of $N_D+1$ dimers by placing an additional
dimer on a dimer vacancy.  The number of new configurations obtained by
placing a dimer on one of the $N_V$ dimer vacancies
of each configuration of $N_D$ dimers is thus $\overline N_V W_N(N_D)$, where
$\overline N_V$ is the average number of dimer vacancies for configurations
having $N_D$ dimers.  However, we note that in this new set of configurations
each configuration occurs $N_D+1$ times because each dimer in the new configuration
could have been the one newly added.  So\cite{FN1}
\begin{eqnarray}
{W_N(N_D+1) \over W_N(N_D)} &=& {\overline N_V \over N_D+1} \ ,
\end{eqnarray}
where we will replace $N_D+1$ by $N_D$ in the thermodynamic limit.  Applying this
for the most probable value of $N_D$ gives the simple result\cite{CC1}
\begin{eqnarray}
{\overline N_V \over \overline N_D} &=& {1 \over z} \ .
\label{NVNDEQ} \end{eqnarray}
If we introduce the density of dimer vacancies $\rho_V$ by $\rho_V =
\overline N_V/N_B$, where $N_B$ is the total number of bonds,
then this may be written as
\begin{eqnarray}
{\rho_V \over \rho} &=& {1 \over z} \ .
\label{RHOVEQ} \end{eqnarray}
Note that $\rho$ (or $\rho_V$) is defined to be the total number
of dimer (or dimer vacancies) divided by the total number of bonds.
These quantities can also be defined locally for a subsystem of bonds.
It is not obvious that when $\rho$ and $\rho_V$ are replaced by their
local versions that Eq. (\ref{RHOVEQ}) will still hold. Indeed, for the Ising
model on a Cayley tree in an applied field, the total magnetization
divided by the total number of sites is not the same as the local magnetization
of a single site near the center of the tree.  We now argue that 
$\rho_V/\rho$ does {\it not} depend on position within the tree.
Consider two subsystems\cite{FN4} S$_1$ and S$_2$ which consist of $N_1$ and $N_2$
bonds, respectively, where $N_1$ and $N_2$ are both large compared to
1 (so that we do not need to worry about discreteness effects).
Suppose $\rho_V/\rho$ assumes {\it different} values within these two
subsystems.  Then, if we move a dimer from subsystem S$_1$ to subsystem
S$_2$, the ratio of the number of configurations $W$ before moving
the dimer to that, $W'$, after moving the dimer is
\begin{eqnarray}
{W \over W'} &=& {\rho_{V1} \rho_2 \over \rho_{V2} \rho_1} \ ,
\end{eqnarray}
where the subscripts identify the subsystem in question.  We may view
the $\rho_{Vn}$ for subsystem S$_n$ as being order parameters.  In order
that $W_N(N_D)$ actually be maximal with respect to variation of these
order parameters, it must be stationary with
respect to moving a dimer.  So $W/W'=1$, which implies that
$\rho_V/\rho$ is the same for all large subsystems.  Thus
Eq. (\ref{RHOVEQ}), which initially involved the ratio of global
properties, is actually valid when interpreted as a relation
between local quantities.

We now express $\rho_V$ as an explicit function of $\rho$ in a
region far from the boundary,
so that Eq. (\ref{RHOVEQ}) yields the equation of state we seek.
If A and B are nearest neighboring sites, then
\begin{eqnarray}
\rho_V &=& P({\rm sites \ A \ and \ B \ are \ vacant}) \nonumber \\
&=& P({\rm A \ is \ vacant}) \nonumber \\ && \ \times
P({\rm B \ is \ vacant, \ given \ that \ A \ is \ vacant}) \nonumber \\
&=& (1-q \rho) \nonumber \\ && \ \times
P({\rm B \ is \ vacant, \ given \ that \ A \ is \ vacant}) \ .
\label{NVEQ} \end{eqnarray}
Here the symbol $P( )$ denotes the probability of the event inside the
parentheses. The basic probability space consists of all configurations
of dimers on the lattice. The (unnormalized) probability assigned to a
configuration is $z^{N_D}$, where $N_D$ is the number of dimers in the
configuration. Equivalently, we could limit our probability space to the
set of configurations containing exactly $N_D$ dimers, with all 
configurations equiprobable. The conditional probability [the last
factor in the last line of Eq. (\ref{NVEQ})] is denoted 
$P({\rm B \ vacant} \ |\ {\rm  A \ vacant})$ and is, by definition, equal to
$P({\rm B \ vacant \ and \ A \ vacant})/P({\rm A \ vacant})$. More generally,
for any two events $E_1$ and $E_2$
\begin{eqnarray}
P(E_2|E_1) = P(E_2 \ {\rm and} \ E_1)/ P(E_1) \ .
\label{CONDITION} \end{eqnarray}
We now invoke the approximation in which we replace the factor
$P({\rm B\ vacant}\ |\ {\rm A\ vacant})$ in Eq. (\ref{NVEQ}) by
$P({\rm B \ vacant}\ |\ {\rm bond \ AB \ unoccupied})$.  This
replacement ignores the possibility that at most one of the other bonds
touching site A might be occupied and thereby might indirectly
affect whether site B is occupied or not.  Of course this possible
inaccuracy can only arise if there is some indirect path from site
A to site B not going through the bond AB.
Since this approximation
(which we will refer to as the {\it tree decoupling}) is exact for
the Bethe lattice it leads to the Bethe approximation. Thus
\begin{eqnarray}
\rho_V &=& (1-q\rho ) \nonumber \\ &\times&
P({\rm B \ is \ vacant} \ | \ {\rm bond \ AB \ is \ unoccupied }) \ .
\label{NV1EQ} \end{eqnarray}
The probability that a particular bond is unoccupied is $1 - \rho$,
so that
\begin{eqnarray}
&& P({\rm B\ vacant}\ |\ {\rm bond \ AB\ unoccupied}) \nonumber \\
&=& {P({\rm B\ vacant\ and\ bond \ AB\ unoccupied}) \over
P({\rm bond\ AB\ unoccupied})} \nonumber \\
&=& {P({\rm B\ vacant\ and\ bond \ AB\ unoccupied}) \over
(1-\rho )} \  .
\label{NV2EQ} \end{eqnarray}
If B is vacant, the bond AB must be unoccupied; therefore the numerator
of Eq. (\ref{NV2EQ}) is just $P({\rm B \ vacant}) =1-q\rho$. Thus
\begin{eqnarray}
\rho_V &=& (1-q\rho )^2/(1-\rho )\ ,
\end{eqnarray}
which, in combination with Eq. (\ref{RHOVEQ}), leads to
Eq. (\ref{BETHE1EQ}).  We have not succeeded in constructing a
simple argument to estimate the magnitude of the difference
between $P({\rm B\ vacant}\ | \ {\rm  A\ vacant})$ and
$P({\rm B\ vacant}\ | \ {\rm bond \ AB\ unoccupied})$ for
$d$-dimensional lattices. Qualitatively,
the important point is that, within the family of configurations in which
bond AB is unoccupied, the probability that site B be vacant is not
significantly influenced by the presence of a dimer on (at most) one of
the other bonds emanating from A. 

In Appendix C we present a simple calculation of the major 
correction to the tree approximation on a planar triangular lattice. 
However, the calculation is not the first term in a systematic series 
and is not easily extended to the square lattice. In the next section 
we will present a formalism which enables one to systematically
generate corrections to the tree approximation. 

\subsection{Solution by Construction of an Effective Hamiltonian} 

We now apply a technique introduced previously\cite{ABH1} in order to
treat here the
nonthermal statistical problem of constructing a generating function
for covering a lattice with 0, 1, 2, $\dots$ hard-core dimers.
We wish to identify this generating function with a partition function,
$Z$ of the form $Z= {\rm Tr} \exp (- \beta {\cal H})$, where ${\cal H}$
can then be interpreted as the Hamiltonian for the statistical problem.

The first step in this program is obviously to construct the effective
Hamiltonian.  This can be done by writing\cite{YS}
\begin{eqnarray}
e^{-\beta {\cal H}} &=& e^{\sum_{\langle ij \rangle} z s_i s_j} \ ,
\label{YSH} \end{eqnarray}
where $\langle ij \rangle$ indicates that the sum is over pairs
of nearest neighbors.  For the expansion of the partition function in
powers of $z$ to count all possible dimer configurations, where $z$
is the dimer activity.  the following {\it trace rules} are imposed
on the operators $s_i$:
\begin{eqnarray}
{\rm Tr}_i s_i^n = C_n \hspace{0.5 in} n=0, 1, \dots \ ,
\label{RULES} \end{eqnarray}
where ${\rm Tr}_i$ indicates a trace over states of site $i$
and we set $C_0=C_1=1$ and $C_n=0$ for $n >1$.  It
is not actually necessary to explicitly construct such an
operator because the only property of these operators we need in
order to construct the partition function is the trace rules
of Eq. (\ref{RULES}).  We see that the fact that the trace of
two or more operators at the same site vanishes, implements
exactly the hard-core constraint for dimers.  Thus the partition
function $Z= {\rm Tr} \exp (- \beta {\cal H})$ will indeed
give the grand partition function for dimers as a function of
their chemical potential $\mu$:
\begin{eqnarray}
Z &=& \sum_{\cal C} e^{\beta \mu n({\cal C})} \ ,
\end{eqnarray}
where the sum is over all configurations ${\cal C}$ of
dimers and $n({\cal C})$ is the number of dimers present in
the configuration ${\cal C}$.  From $Z$ we can get the
fraction of sites covered by dimers $\rho$ as a function of the
dimer chemical potential via
\begin{eqnarray}
\rho & \equiv & {2 \overline N_D \over Nq}
= {2 \over Nq \beta} {\partial \ln Z \over \partial \mu } =
{2 \over Nq} {d \ln Z \over d \ln z} \ .
\label{DIMER} \end{eqnarray}
Since the ``spin operators'' $s_i$ commute with one another,
we have a mapping of the athermal problem of dimers on a lattice
into a statistical mechanical problem involving classical
spins with a given Hamiltonian.

Now we use this mapping to a) construct the exact solution for
the partition function for a Bethe lattice and b) generate series
expansions for finite dimensional lattices.  To do that we develop
a perturbation theory for
a periodic lattice in which the leading term contains the sum
of all contributions from tree diagrams.  For this purpose we
write $Z = {\rm Tr} \prod_{\langle ij \rangle} f_{ij}$, where
$f_{ij} = 1+ z s_i s_j$ [the trace rules allow us to linearize the
exponential in Eq. (\ref{YSH})] or equivalently
\begin{eqnarray}
Z &=& {\rm Tr} \Biggl\{ \left[ \prod_i g_i^q \right] \left[\prod_{\langle ij \rangle}
{f_{ij} \over g_i g_j } \right] \Biggr\} \ ,
\end{eqnarray}
where $g$ can be chosen arbitrarily. We evaluate this perturbatively as
\begin{eqnarray}
Z &=& {\rm Tr} \Biggl\{ \left[ \prod_i g_i^q \right] \left[ \prod_{\langle ij \rangle}
1 + \lambda V_{ij} \right] \Biggr\} \ ,
\end{eqnarray}
where $V_{ij} = f_{ij}/(g_ig_j) - 1 $.  We expand in powers of $\lambda$
which we set equal to unity at the end.  Each term in this expansion
which involves at least one power of $\lambda$ can be associated with
a diagram in which the factor $V_{ij}$ is
associated with a line connecting sites $i$ and $j$.
We now choose $g$ so that diagrams having at least one line which
is connected to only a single site ($j$) give zero contribution. 
Since all diagrams on a Bethe lattice have at least one free end,
this choice of $g$ will lead to an exact evaluation of the partition
function for a Bethe lattice and will enable us to generate loop corrections
for $d$-dimensional lattices.  The condition we implement is that
\begin{eqnarray}
{\rm Tr}_j \left( g_j^q V_{ij} \right) = 0 \ ,
\end{eqnarray}
which can be written in the form
\begin{eqnarray}
g_i = { {\rm Tr}_j \left( f_{ij} g_j^\sigma \right) \over {\rm Tr}_j g_j^q} \ ,
\label{NONLIN}
\end{eqnarray}
where $\sigma=q-1$.  This is a nonlinear equation for the
function $g_j$, but in view of the trace rules it is easily solved. 
From the form of $f_{ij}$ one sees that $g_i$ has to be of the form
\begin{eqnarray}
g_i = A + B s_i \ .
\end{eqnarray}
By substituting this form into Eq. (\ref{NONLIN}):
\begin{eqnarray}
A + B s_i = { {\rm Tr}_j \left[ (1+ z s_i s_j)
(A+B s_j)^\sigma \right] \over {\rm Tr}_j (A+B s_j)^q } \ .
\end{eqnarray}
Using the trace rules we rewrite the right-hand side of
this equation so that
\begin{eqnarray}
A + B s_i = { A^\sigma + \sigma A^{\sigma-1} B 
+ s_i (zA^\sigma ) \over A^q + qA^\sigma B } \ .
\end{eqnarray}
This gives rise to the two equations
\begin{eqnarray}
A &=& {A^\sigma + \sigma A^{\sigma-1} B \over A^q +q A^\sigma B} =
{A + \sigma B \over A^2 +qAB } \label{EQNA} \\
B &=& {zA^\sigma \over A^q +q A^\sigma B} = {z \over A+qB} \ .
\label{EQNB} \end{eqnarray}
We may solve Eq. (\ref{EQNA}) for $B$ as
\begin{eqnarray}
B &=& {A^3 - A \over \sigma - q A^2 } \ .
\end{eqnarray}
Substituting this into Eq. (\ref{EQNB}) leads to
\begin{eqnarray}
A^4 \left( 1 + q^2 z \right) - A^2 \left( 2q \sigma z +
1 \right) + z \sigma^2 = 0 \ .
\end{eqnarray}
Thus
\begin{eqnarray}
A^2 &=& {2 qz\sigma +1 + \sqrt{1 + 4 z \sigma} \over
2(1+q^2 z) } \ .
\label{ACAYEQ} \end{eqnarray}
(We chose the positive sign before the square root to ensure that
$A \rightarrow 1$ as $z \rightarrow 0$.) 
Then for the Bethe lattice we have the exact result
\begin{eqnarray}
Z^{1/N} &=& {\rm Tr}_i g_i^q ={\rm Tr}_i [A + B s_i]^q \nonumber \\
&=& A^q + qBA^\sigma \ . 
\end{eqnarray}
After some algebra it can be shown that when this result is inserted into
Eq. (\ref{DIMER}) we recover the result of Eq. (\ref{BETHE2EQ}).
 
Unfortunately, this formalism leads to an expansion of the partition
function $Z$, whereas for $d$ dimensional lattices, we would prefer to
have an expansion for the free energy per site, $F=(1/N)\ln Z$.
For that purpose we consider an expansion of the quantity $Z^n$,
which in the limit $n \rightarrow 0$ is $1+n \ln Z = 1 + nNF$.  To
obtain $Z^n$ we introduce the $n$-replicated Hamiltonian
\begin{eqnarray}
e^{- \beta {\cal H}_n} &=& \prod_{\alpha=1}^n \prod_{\langle ij \rangle}
[1 + z s_{i\alpha} s_{j\alpha}] \ ,
\end{eqnarray}
where $s_{i\alpha}$ and $s_{i \beta}$ are independent operators
for $\alpha \not=\beta$ and for each replica index $s_{i\alpha}$
obeys the same trace rules as in Eq. (\ref{RULES}).  Usually replicas
are introduced to perform the quenched average (over $\ln Z$) for
random problems in which case the averaging leads to interactions
between different replicas.  Here we introduce replicas simply to
facilitate construction of an expansion of $\ln Z$ and there are
no interactions between different replicas.  The partition
function, $Z_{\rm rep}$ associated with the replicated Hamiltonian is
\begin{eqnarray}
Z_{\rm rep} &\equiv& {\rm Tr} e^{-\beta {\cal H}_n} = Z^n \ .
\end{eqnarray}

Thus we solve Eq. (\ref{NONLIN}) with $f_{ij} = \prod_\alpha (1+zs_{i\alpha}
s_{j\alpha})$.  Because different replicas are independent of one
another, the solution to Eq. (\ref{NONLIN}) is of the form
\begin{eqnarray}
g_i = C \prod_{\alpha=1}^n [1 + D s_{i\alpha}] \equiv C \hat g_i \ .
\end{eqnarray}
Because we need the partition function $Z_{\rm rep}$ to order $n$,\cite{FN}
we must evaluate $C$ up to linear order in  $n$ but $D$ can be
evaluated for $n=0$ because it always
appears in connection with a sum over replica indices which give a factor
of $n$.  The terms in $g_i$ in Eq. (\ref{NONLIN}) independent of $s_{i\alpha}$ give
\begin{eqnarray}
C = {\rm Tr}_j C^\sigma \hat g_j^\sigma / {\rm Tr}_j C^q \hat g_j^q
\end{eqnarray}
or
\begin{eqnarray}
C^2 &=& {\rm Tr}_j \hat g_j^\sigma / {\rm Tr}_j \hat g_j^q \nonumber \\
&=& (1+\sigma D)^n / (1+qD)^n \ ,
\end{eqnarray}
which to linear order in  $n$ gives
\begin{eqnarray}
C &=& 1 + {n \over 2} \ln \left( {1+\sigma D \over 1 + qD} \right) \ .
\end{eqnarray}
The terms in $g_i$ in Eq. (\ref{NONLIN}) linear in $s_{i\alpha}$ give
\begin{eqnarray}
CD = {z {\rm Tr}_j C^\sigma s_{j\alpha} \prod_\beta (1+D s_{j\beta})^\sigma
\over {\rm Tr}_j C^q \prod_\beta (1+D s_{j\beta})^q} \ .
\end{eqnarray}
For $n=0$ this is
\begin{eqnarray}
D &=& z (1+\sigma D)^{-1} 
\label{DEQ} \end{eqnarray}
which gives
\begin{eqnarray}
D &=& {-1 + r \over 2 \sigma} \ ,
\end{eqnarray}
where $r\equiv \sqrt{1+4\sigma z}$.  

Then the first term in the expansion of $F$ is
\begin{eqnarray}
F &=& {d \over dn} \Biggl[ {\rm Tr}_j (C \hat g_j)^q \Biggr]_{n=0} 
\nonumber \\ &=& {q \over 2} \ln \left( {1+\sigma D \over 1 + q D} \right) +
\ln (1+qD) \ . 
\label{FREEEQ} \end{eqnarray}
After some algebra one can show that when this is substituted into
Eq. (\ref{DIMER}) we recover Eq. (2).

\subsection{Loop Corrections}

Here we consider the expansion in powers of $V_{ij} \equiv f_{ij}/(g_ig_j)-1$.
Our first objective is to show that contributions to $Z_{\rm rep}$ from
disconnected diagrams are of order $n^2$ or higher and hence can be dropped.
To see this we write
\begin{eqnarray}
V_{ij} &=& C^{-2} \prod_\alpha \Biggl( [1 + zs_{i\alpha } s_{j \alpha}] 
[1 - D s_{i \alpha} ] [1 - D s_{j \alpha}] \Biggr) \nonumber \\ && \ - 1 \ .
\end{eqnarray}
The contribution to $Z_{\rm rep}$ from a disconnected diagram
is simply the product of the contributions from each connected component.
We now argue that the contribution to $Z_{\rm rep}$ from a single
connected diagram is of order $n$.
Note that $V_{ij}$ is a multinomial in the $s_{\alpha}$'s whose constant term
is proportional to $n$.  Thus the contribution to $Z_{\rm rep}$
will get at least one factor of $n$, either from the constant term in
a $V_{ij}$, or from a sum over replica indices from terms in
a $V_{ij}$ involving an $s$ operator. Thus using the $n\rightarrow 0$
limit of the replica formalism we have eliminated unlinked
diagrams. 

\begin{figure}
\psfig{figure=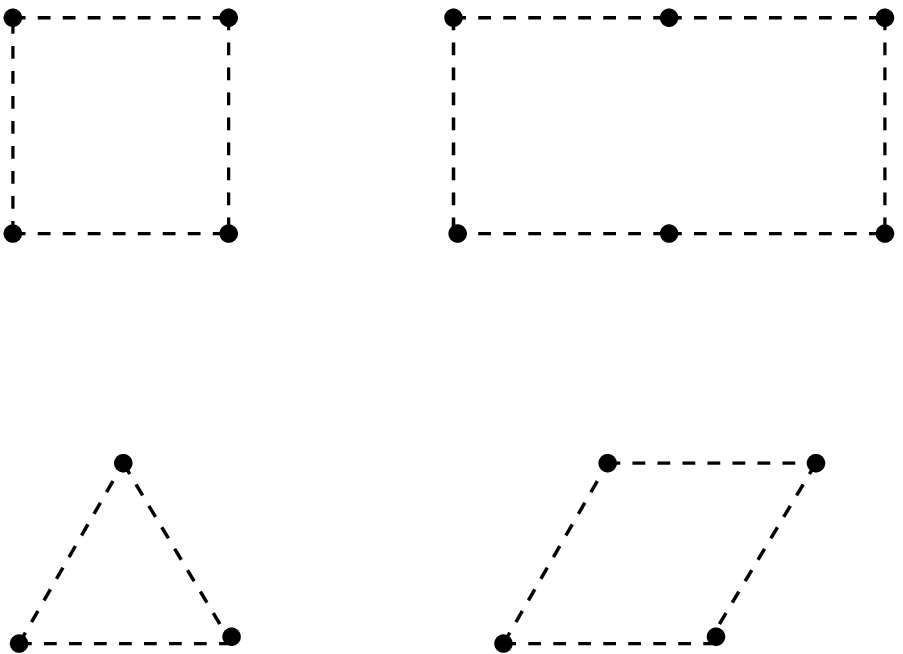}
\caption{Leading loop corrections to $Z^n$ for the hypercubic lattice
(top) and for the plane triangular lattice (bottom).}
\label{LOOP}
\end{figure}

Now we consider the leading loop corrections to $Z^n$.   These
come from the smallest loops that can be drawn on the lattice.
In Fig. \ref{LOOP} we show these loops for both the $d$ dimensional
hypercubic lattice and the plane triangular lattice.  Here we will
explicitly evaluate only the corrections from the smallest possible
loop diagrams. 

We will now show that this expansion involves evaluating a
modified dimer partition function for diagrams with no free ends.
The contribution to the replica partition function from
such a diagram $\Gamma$ is given by
\begin{eqnarray}
\delta Z_{\rm rep}(\Gamma) &=& {\rm Tr} \Biggl\{
\prod_{i \in \Gamma} g_i^q \prod_{\langle ij \rangle \in \Gamma}
\left( {f_{ij} \over g_i g_j} -1 \right) \Biggr\} \nonumber \\
&=& n \delta F(\Gamma) + {\cal O} (n^2) \ .
\end{eqnarray}
The notation $i \in \Gamma$ means that the site $i$ is a site covered by
at least one bond of $\Gamma$.  We do not change the result for
$\delta F(\Gamma)$ if we divide this by a quantity which differs from
unity by terms of order $n$. So, for later convenience we write
\begin{eqnarray}
\delta Z_{\rm rep} (\Gamma) &=& { {\rm Tr} \Biggl\{
\prod_{i \in \Gamma} g_i^q \prod_{\langle ij \rangle \in \Gamma}
\left( {f_{ij} \over g_i g_j} -1 \right) \Biggr\} \over
{\rm Tr} \prod_{i \in \Gamma} g_i^q } \ .
\end{eqnarray}
We now expand the product over bonds, into its $2^{N_B(\Gamma)}$ terms, where
$N_B(\Gamma)$ is the number of bonds in the set $\Gamma$.  In so doing
note that when considered as a multinomial series in
$\{s_{i \alpha}\}$, each term has the limiting value unity
as $n \rightarrow 0$ (but the sum of all $2^{N_B}$ terms is zero in this limit).
Since we are interested in the limit $n \rightarrow 0$, we consider
$d \delta Z_{\rm rep}(\Gamma)/dn$ (evaluated at $n=0$) and write
\begin{eqnarray}
{d\delta Z_{\rm rep}(\Gamma) \over dn} &=& \sum_{\gamma \in \Gamma}
(-1)^{N_B(\Gamma)-N_B(\gamma)} G(\gamma) \ ,
\label{CUM1EQ} \end{eqnarray}
where the sum is over the $2^{N_B(\Gamma)}-1$ nonempty subsets
$\gamma$ of $\Gamma$, including $\gamma=\Gamma$. (The
term corresponding to the empty set is unity and therefore drops
out when differentiated with respect to $n$.)  Here
\begin{eqnarray}
G(\gamma) &=& {d \over dn} \Biggl[ {{\rm Tr} \prod_{i \in \gamma}
g_i^q \prod_{\langle ij \rangle \in \gamma} {f_{ij} \over g_i g_j}
\over {\rm Tr} \prod_{i \in \gamma} g_i^q } \Biggr] \ .
\end{eqnarray}
Because we divided by the factor in the denominator, it is no longer
necessary to involve sites in $\Gamma$ which are not in $\gamma$.
The subtractions of subdiagrams indicated in Eq. (\ref{CUM1EQ}) 
defines the cumulant operation (indicated by a subscript ``c''),
so we write
\begin{eqnarray}
\delta F(\Gamma) &=& G_c(\Gamma) \ ,
\end{eqnarray}
where $G_c(\Gamma)$ is the right-hand side of Eq. (\ref{CUM1EQ}).
This representation is not very efficient because it contains
$2^{N_B(\Gamma)}-1$ terms, many of which are either zero or
are identical to one another.  The following equivalent recursive
definition is more convenient:
\begin{eqnarray}
G_c(\gamma) = G(\gamma) - {\sum_{\gamma' \in \gamma}}^\prime
G_c(\gamma^\prime) \ ,
\label{CUM2EQ} \end{eqnarray}
where the prime indicates that in the sum over subsets $\gamma^\prime$
we do not include $\gamma^\prime=\gamma$.  For the smallest loop on
a hypercubic lattice, namely a square of four bonds, there are no nonzero
subtractions and $G_c(\gamma)$ is equal to its ``bare'' value $G(\gamma )$.
More generally Eq. (\ref{CUM2EQ}) has many fewer terms than Eq. (\ref{CUM1EQ})
and furthermore, all the cumulants of the subdiagrams will have been
previously calculated in a lower order calculation.  So, to implement
the cumulant subtraction we only need to subtract the cumulant
contributions of subgraphs with no free ends.

It remains to discuss the calculation of $G(\gamma)$.  We take
\begin{eqnarray}
g_i = C \prod_\alpha (1+Ds_{i\alpha}) \equiv C \prod_\alpha
\hat g_{i \alpha}
\end{eqnarray}
and
\begin{eqnarray}
f_{ij} &=& \prod_\alpha (1+z s_{i\alpha} s_{j \alpha}) \equiv \prod_\alpha f_{ij;\alpha} \ .
\end{eqnarray}
Then
\begin{eqnarray}
G(\gamma) &=& {d \over dn} \Biggl[ { C^{-2N_B(\gamma)}
\prod_\alpha {\rm Tr} \prod_{i \in \gamma} \hat g_{i\alpha}^q 
\prod_{\langle ij \rangle \in \gamma} {f_{ij;\alpha} \over \hat g_{i \alpha}
\hat g_{j\alpha}} \over \prod_\alpha {\rm Tr} \prod_{i \in \gamma} \hat g_{i \alpha}^q } 
\Biggr] \nonumber \\ &=& {d \over dn} \Biggl[ C^{-2N_B(\gamma)}
(1+qD)^{-N_s(\gamma) n} Q(\gamma)^n \Biggr] \ ,
\end{eqnarray}
where $N_s(\gamma)$ is the number of sites in $\gamma$ and $Q(\gamma)$
is a partition function for the graph $\gamma$:
\begin{eqnarray}
Q(\gamma) &=& {\rm Tr} \Biggl[ \prod_{i \in \gamma} (1+Ds_i)^q
\prod_{\langle ij \rangle \in \gamma}
{1 + z s_i s_j \over (1+Ds_i)(1+Ds_j)} \Biggr] \nonumber \\
&=& {\rm Tr} \Biggl[ \prod_{i \in \gamma} (1+Ds_i)^{q-q_i(\gamma)}
\prod_{\langle ij \rangle \in \gamma}
(1 + z s_i s_j ) \Biggr] \ ,
\end{eqnarray}
where $q_i(\gamma)$ is the number of sites neighboring to $i$
which are connected to $i$ by a bond in $\gamma$.
Because of the trace rules, the product over sites mimics a site-dependent
monomer activity $z_i\equiv 1+[q-q_i(\gamma)]D$ and we therefore have
\begin{eqnarray}
Q(\gamma) &=& \Biggl[ \prod_{i \in \gamma} z_i \Biggr]
\hat Q(\gamma ; z_{ij}= z/(z_iz_j)) \ ,
\end{eqnarray}
where $\hat Q$ is the grand partition function for the set of bonds
$\gamma$ in which the bond $\langle ij \rangle$ has the bond-dependent
activity $z_{ij} \equiv z/(z_iz_j)$:
\begin{eqnarray}
\hat Q(\gamma ; \{z_{ij}\}) &=&
{\rm Tr} \prod_{\langle ij \rangle \in \gamma } (1 + z_{ij} s_i s_j) \ .
\end{eqnarray}
Thus the renormalized free energy associated with a
diagram with no free ends is only slightly more complicated than
its unrenormalized ($g_i=1$) value.  To summarize:
\begin{eqnarray}
G(\gamma) &=& - N_s(\gamma) \ln (1+qD)
+ N_B(\gamma) \ln \left( {1 + q D \over 1 + \sigma D} \right) \nonumber \\
&+& \sum_{i \in \gamma} \ln \left( 1 + [q-q_i(\gamma)]D \right)
+ \ln \hat Q(\gamma ; \{ z_{ij} \} ) \ .
\label{GEQ} \end{eqnarray}
It is a remarkable fact that for a diagram $\gamma$ with no loops,
$G(\gamma)$ vanishes and this forms a nice check of computer
programs used to evaluate $G(\gamma)$ for an arbitrary diagram.
(This is easy to check for small diagrams.)  Furthermore, for a 
diagram with loops and which has a free end, $G(\gamma)$ does
{\it not} vanish, but its cumulant $G_c(\gamma)$ {\it does} vanish.
(This is also a nice check of computer programs.)

For the hypercubic lattice we consider the leading correction from
a square of four nearest neighbor bonds, $\gamma$. So we use Eq. (\ref{GEQ})
with $q=4$ and $q_i(\gamma)=2$, so that $z_{ij}=z/(1+2D)^2$.  Then
$\hat Q(\gamma)= 1 + 4 z_{ij} + 2 z_{ij}^2$ and
\begin{eqnarray}
G(\gamma ) &=& 4 \ln (1+2D) - 4 \ln (1+4D) + 4 \ln \left(
{1 + 4D \over 1 + 3D} \right) \nonumber \\
&& + \ln \Biggl[ 1 + {4z \over (1+2D)^2} + {2z^2 \over (1+2D)^4} \Biggr] \nonumber \\
&=& - 4 \ln (1+3D) \nonumber \\ && \ 
+ \ln \left[ (1+2D)^4 +4z(1+2D)^2 + 2z^2 \right] \ .
\end{eqnarray}
Now use $1+3D=z/D$ from Eq. (\ref{DEQ}), so that
\begin{eqnarray}
G(\gamma) &=& \ln \Biggl[ \left( 1 - {D^2 \over z} \right)^4 + 4 {D^2 \over z}
\left( 1 - {D^2 \over z} \right)^2 + 2 {D^4 \over z^2} \Biggr] \nonumber \\
&=& \ln \Biggl[ 1 + {D^8 \over z^4} \Biggr] \ .
\end{eqnarray}
For $z$ small $D$ is proportional to $z$ and this diagram gives a contribution
to the free energy of order $z^4$. Since there are $d(d-1)/2$ squares
per site, the perturbative contribution to the free energy per site is
\begin{eqnarray}
\delta F &=& {1 \over 2} d(d-1) \ln \left[  1 + {D^8 \over z^4} \right] \ . 
\label{LOOPF} \end{eqnarray}
The dimer density then follows using Eq. (\ref{DIMER}) and the results
are given in Table I for a square lattice.

\begin{table}
\caption{First Loop Correction for the Square Lattice
for the density of dimers ($\pi \equiv \overline N_D/N$)
as a function of $z$.  Here $\pi_0$ is the value
for the Bethe lattice for $q=4$, $\pi_1$ is the value
when the first loop correction, Eq. (\protect\ref{LOOPF}), 
is included, and $\pi_{\rm E}$ is
the ``exact'' result obtained by extrapolation\protect\cite{CC1}
of the series of Gaunt.\protect\cite{GAUNT}}

\vspace{0.15 in}
\begin{tabular} {| c c c c |} \hline
$z$ & $\pi_0$ & $\pi_1$ & $\pi_{\rm E}$ \\ \hline
\ \ 0.005180\ \ & \ \  0.010000\ \  &\ \ 0.010000\ \ &\ \ 0.010000\ \ \\ 
0.010742&     0.020000 &     0.020000 &     0.020000 \\ 
0.023156 &     0.039999 &     0.039999 &     0.040000 \\ 
0.037575 &     0.059997 &     0.060000 &     0.060000 \\ 
0.054413 &     0.079990 &     0.080000 &     0.080000 \\ 
0.074195 &     0.099978 &     0.100000 &     0.100000 \\ 
0.097587 &     0.119954 &     0.119999 &     0.120000 \\ 
0.125452 &     0.139917 &     0.139998 &     0.140000 \\ 
0.158915 &     0.159862 &     0.159996 &     0.160000 \\ 
0.199466 &     0.179784 &     0.179991 &     0.180000 \\ 
0.249111 &     0.199679 &     0.199982 &     0.200000 \\ 
0.310597 &     0.219543 &     0.219968 &     0.220000 \\ 
0.387766 &     0.239370 &     0.239945 &     0.240000 \\ 
0.486115 &     0.259158 &     0.259909 &     0.260000 \\ 
0.613728 &     0.278904 &     0.279857 &     0.280000 \\ 
0.782881 &     0.298607 &     0.299783 &     0.300000 \\ 
1.012941 &     0.318271 &     0.319684 &     0.320000 \\ 
1.335914 &     0.337903 &     0.339556 &     0.340000 \\ 
1.807776 &     0.357516 &     0.359397 &     0.360000 \\ 
2.533661 &     0.377128 &     0.379204 &     0.380000 \\ 
3.730072 &     0.396761 &     0.398970 &     0.400000 \\ 
5.902289 &     0.416436 &     0.418682 &     0.420000 \\ 
10.466788 &     0.436180 &     0.438318 &     0.440000 \\ 
22.802101 &     0.456068 &     0.457892 &     0.460000 \\ 
81.673705 &     0.476432 &     0.477641 &     0.480000 \\ 
\hline \end{tabular} \end{table}
 
We now identify the expansion parameters in this formulation.  It is clear that
the free energy, $F$, is obtained as a sum of contributions associated
with diagrams having no free ends, the smallest of which are shown
in Fig. \ref{LOOP}.  This development will lead to an evaluation of $F$
as a power series in the activity:
\begin{eqnarray}
F &=& \sum_n F_n z^n \ .
\end{eqnarray}
From Eq. (\ref{FREEEQ}) we see that for the Bethe lattice and for large $q$
$F_n \sim q^n/n$.  We assert that the contributions to $F_n$ from a diagram
$\gamma$ [which we denote $\delta F(\gamma )$] are of order
\begin{eqnarray}
\delta F_n(\gamma ) / F_n \sim q^{-r(\gamma )} \ ,
\end{eqnarray}
where $r(\gamma )$ is an integer which increases with the size and complexity
of the diagram. For instance, for a square of four bonds Eq. (\ref{LOOPF})
indicates that $r=2$ and for a loop of $2n$ bonds a similar result shows that
$r=n$.  So our diagrammatic formulation generates corrections in inverse powers
of $q$.  Furthermore, if one expands $f_{ij}$ and $g_i$ in powers of $z$, one
sees that $V_{ij}$ is of order $z$.  This means that a diagram $\gamma$ with
$N_B(\gamma )$ bonds contributes to $F(q)$ at order $z^{N_B(\gamma )}$
{\it and \ higher}.  So our formulation involves the two expansion parameters
$z$ and $1/q$.

An entirely analogous calculation gives the leading loop correction for
the triangular lattice from triangles as
\begin{eqnarray}
\delta F = N_T \ln \left[ 1 - {D^6 \over z^3} \right] \ ,
\label{TRIEQ} \end{eqnarray}
where $N_T=2$ is the number of triangles per site.  Thus we see that the
leading correction to the Bethe lattice result has the opposite sign for
triangular lattices as compared to hypercubic lattices. The results
based on Eq. (\ref{TRIEQ}) are given in Table II.

\begin{table}
\caption{First Loop Correction for the Triangular Lattice.
The notation is as in Table I, but where $\pi_1$ includes the
loop correction of Eq. (\protect\ref{TRIEQ}) and $\pi_{\rm G}$
includes the loop correction derived geometrically in Appendix C.}

\vspace{0.15 in}
\begin{tabular} {| c c c c c |} \hline
$z$ & $\pi_0$ & $\pi_1$ & $\pi_{\rm E}$ & $\pi_{\rm G}$ \\ \hline
\ \ 0.009999 \ \ & \ \  0.009999\ \ &\ \ 0.009999\ \ &\ \ 0.010000\ \
&0.003459\ \ \\ 
0.011206 &     0.030004 &     0.029999 &     0.030000 & 0.029999\\ 
0.025331 &     0.060040 &     0.059998 &     0.060000 & 0.059997\\ 
0.043360 &     0.090121 &     0.089991 &     0.090000 & 0.089987\\ 
0.066712 &     0.120261 &     0.119975 &     0.120000 & 0.119962\\ 
0.097473 &     0.150457 &     0.149943 &     0.150000 &0.149912\\ 
0.138811 &     0.180705 &     0.179892 &     0.180000 & 0.179831\\ 
0.195705 &     0.210994 &     0.209822 &     0.210000 & 0.209712\\ 
0.276329 &     0.241312 &     0.239741 &     0.240000 & 0.239559\\ 
0.394826 &     0.271650 &     0.269667 &     0.270000 & 0.269386\\ 
0.577329 &     0.302001 &     0.299629 &     0.300000 & 0.299221\\ 
0.876264 &     0.332353 &     0.329663 &     0.330000 & 0.329101\\ 
1.408720 &     0.362668 &     0.359786 &     0.360000 & 0.359056\\ 
2.477977 &     0.392849 &     0.389972 &     0.390000 & 0.389084\\ 
5.067826 &     0.422724 &     0.420118 &     0.420000 & 0.419135\\ 
13.932546 &    0.452078 &     0.450085 &     0.450000 & 0.449161\\ 
91.850988 &    0.480860 &     0.479893 &     0.480000 & 0.479338\\ 
\hline \end{tabular} \end{table}

The present development appears to be related to that of Ref. \onlinecite{JFN},
but the detailed relationship of the two approaches is not clear to us.  We
note that the disconnected diagram of Fig. 3d of Nagle\cite{JFN} (which
``is required'' for his dimer series) does not appear in our approach. 

\subsection{ANISOTROPY}

We may generalize the above model to allow for different activities
along different coordinate axes.  For that purpose we relate the
square lattice to a Bethe lattice in which each site is surrounded
by four bonds, two of which we arbitrarily label as ``horizontal''
(or $x$) bonds and the others as ``vertical'' (or $y$) bonds.  Dimers
along the $\alpha$ axis have activity $z_\alpha$ for $\alpha=x,y$.
The fraction of $\alpha$ bonds covered by dimers will be denoted
$\rho_\alpha$ and the fraction of $\alpha$ bonds which are dimer
vacancies will be denoted $\rho_{V\alpha}$

\subsubsection{Geometrical Approach}

The geometrical reasoning used before yields
\begin{eqnarray}
z_\alpha^{-1} &=& \rho_{V\alpha} / \rho_\alpha \ .
\end{eqnarray}
The tree approximation, which is exact for the Bethe lattice, then
yields
\begin{eqnarray}
\rho_{V\alpha} = W/(1- \rho_\alpha ) \ ,
\end{eqnarray}
where $W^{1/2}$ is the probability that a given site is vacant, where
\begin{eqnarray}
W &=& (1- 2 \rho_x - 2 \rho_y)^2\ ,
\end{eqnarray}
so that
\begin{eqnarray}
{\rho_x (1- \rho_x) \over z_x} &=& {\rho_y (1- \rho_y) \over z_y} 
= W \ .
\label{RHOGEQ} \end{eqnarray}
Thus
\begin{eqnarray}
\rho_\alpha = [1 - \sqrt{1-4z_\alpha W}]/2 
\end{eqnarray}
and
\begin{eqnarray}
W &=& \Biggl( \sqrt{1-4z_xW} + \sqrt{1-4z_yW} -1 \Biggr)^2 \ .
\end{eqnarray}
This equation allows one to calculate $W$ as a function of
$z_x$ and $z_y$, from which all the other relevant quantities
can be obtained.

\subsubsection{Algebraic Approach}

We use the Hamiltonian
\begin{eqnarray}
e^{-\beta {\cal H}} &=&
\prod_{\langle ij \rangle \in H} [1 + z_x s_i s_j]
\prod_{\langle ij \rangle \in V} [1 + z_y s_i s_j]
\end{eqnarray}
where the sum ``$\in$ H'' means we sum over horizontal ($x$)
bonds and similarly for ``$\in$ V.''
The operators $s_i$ obey the same trace rules as before.
We write the partition function as
\begin{eqnarray}
Z &=& {\rm Tr} \prod_i g_{ix}^2 g_{iy}^2 \prod_{\langle ij \rangle}
[1 + V_{ij}] \ ,
\end{eqnarray}
where for horizontal bonds
\begin{eqnarray}
V_{ij} = [1+z_x s_i s_j]/g_{ix}g_{jx} - 1
\end{eqnarray}
and for vertical bonds
\begin{eqnarray}
V_{ij} = [1+z_y s_i s_j]/g_{iy}g_{jy} - 1  \ .
\end{eqnarray}
Now we expand in powers of $V_{ij}$ and require that diagrams
with either a vertical or a horizontal free end vanish.
For horizontal bonds we require that
\begin{eqnarray}
{\rm Tr}_j g_{jx}^2 g_{jy}^2 V_{ij} &=& 0 \ ,
\end{eqnarray}
so that
\begin{eqnarray}
g_{ix} &=& {{\rm Tr}_j g_{jx} g_{jy}^2[1+ z_x s_i s_j]  
\over {\rm Tr}_j g_{jx}^2 g_{jy}^2} \ .
\end{eqnarray}
For vertical bonds we require that
\begin{eqnarray}
{\rm Tr}_j g_{jx}^2 g_{jy}^2 V_{ij} &=& 0 \ ,
\end{eqnarray}
so that
\begin{eqnarray}
g_{ix} &=& {{\rm Tr}_j g_{jx}^2 g_{jy}[1+ z_y s_i s_j]  
\over {\rm Tr}_j g_{jx}^2 g_{jy}^2} \ .
\end{eqnarray}
These equations have a solution of the form
\begin{eqnarray}
g_{i\alpha}= A_\alpha + B_\alpha s_i \ .
\end{eqnarray}
Thus
\begin{eqnarray}
A_x &+& B_x s_i = {{\rm Tr}_j [1+z_x s_i s_j]
[A_x+B_xs_j] [A_y+B_ys_j]^2 \over
{\rm Tr}_j [A_x+B_xs_j]^2 [A_y+B_ys_j]^2} \nonumber \\
&=& {A_xA_y^2 + B_xA_y^2 + 2A_xA_yB_y + s_iz_xA_xA_y^2 \over
A_x^2A_y^2 + 2A_xB_xA_y^2 + 2A_x^2A_yB_y}
\end{eqnarray}
and
\begin{eqnarray}
A_y &+& B_y s_i = {{\rm Tr}_j [1+z_y s_i s_j]
[A_x+B_xs_j]^2 [A_y+B_ys_j] \over
{\rm Tr}_j [A_x+B_xs_j]^2 [A_y+B_ys_j]^2} \nonumber \\
&=& {A_x^2A_y + 2 A_xB_xA_y + A_x^2B_y + s_i z_y A_x^2 A_y \over
A_x^2A_y^2 + 2A_xB_xA_y^2 + 2A_x^2A_yB_y} \ .
\end{eqnarray}
Thus if we set $B_x=b_xA_x$ and $B_y=A_yb_y$, then
\begin{eqnarray}
A_x^2 &=& {1 + b_x + 2 b_y \over 1 + 2 b_x + 2 b_y} \nonumber \\
A_y^2 &=& {1 + 2b_x +  b_y \over 1 + 2 b_x + 2 b_y} \nonumber \\
A_x^2 b_x &=& {z_x \over 1 + 2 b_x + 2 b_y} \nonumber \\
A_y^2 b_y &=& {z_y \over 1 + 2 b_x + 2 b_y} \ .
\end{eqnarray}
Thus
\begin{eqnarray}
b_x &=& {z_x \over 1 + b_x + 2 b_y} \nonumber \\
b_y &=& {z_y \over 1 + 2 b_x + b_y} \ .
\label{BEQ} \end{eqnarray}
These equations reproduce those from the geometrical approach
if one makes the identification 
\begin{eqnarray}
\rho_\alpha = b_\alpha / [1+2b_x+2b_y]
\label{RHOEQ} \end{eqnarray}
and
\begin{eqnarray}
W = (1+2b_x+2b_y)^{-2} \ .
\end{eqnarray}
The density of dimer bonds along $\alpha$ is given by
\begin{eqnarray}
\rho_\alpha &=& {1 \over N} {\partial Z \over \partial z_\alpha} z_\alpha
\nonumber \\ &=& z_\alpha {\partial \over \partial z_\alpha}
\ln \Biggl( A_x^2 A_y^2 [1+2b_x + 2 b_y] \Biggr) \ .
\end{eqnarray}
This can be shown to be equivalent to Eq. (\ref{RHOGEQ}).
Thus we conclude that the algebraic approach agrees with
the much simpler geometrical approach.  But, in principle,
the algebraic approach can be used to generate corrections to
the tree approximation for $d$ dimensional lattices.

\section{INTERACTING DIMERS}

\subsection{Geometrical Derivation of The Exact Result}

Here we consider the case when two dimers separated by  a single
bond (an interacting bond as shown in Fig. 1) have an
interaction energy $- \alpha$.  Thus the energy of a configuration of
dimers on the lattice is  $-N_I \alpha$, where $N_I$ is the
number of interacting bonds.  We start by expressing $N_I$ in terms
of $N_V$.  For that purpose we record the following sum rules
for periodic lattices.
The first sum rule expresses the fact that each bond is uniquely
a member of one of the four sets shown in Fig. 1, so that the
total number of bonds of the lattice $N_B$ is given by
\begin{eqnarray}
N_V + N_I + N_D + N_H &=& N_B = Nq/2 \ ,
\end{eqnarray}
where $N_H$ is the number of half bonds (see Fig. 1). The second sum
rule is obtained by imagining putting a stick on each of the $q$
bonds emanating from each occupied site.  The total number of sticks
is obviously $q(2N_D)$.  Each half bond has one stick, whereas each
interacting bond and each dimer bond is covered by two sticks, so that
\begin{eqnarray}
2qN_D &=& N_H + 2 N_I + 2N_D \ .
\end{eqnarray}
Combining these two relations we get
\begin{eqnarray}
N_I &=& -Nq/2 + (2q-1)N_D + N_V \ .
\end{eqnarray}
Thus the probability $P(N_D)$ that there are $N_D$ dimers on the lattice is
\begin{eqnarray}
P(N_D)&=& Q^{-1} \sum_{\cal G} e^{\beta \alpha N_I} z^{N_D} \nonumber \\
&=& {\rm const} \times [ze^{\beta \alpha (2q-1)}]^{N_D}
\sum_{\cal G} e^{\beta \alpha N_V({\cal G})} \ ,
\label{PNDEQ} \end{eqnarray}
where $Q$ is the partition function, the sum over $\cal G$ runs
over all configurations with $N_D$ dimers on the lattice, and 
$N_V({\cal G})$ is the number of dimer vacancies in the graph ${\cal G}$.
Similarly
\begin{eqnarray}
P(N_D+1) = {\rm const.} \times [ze^{\beta \alpha (2q-1)}]^{N_D+1}
\sum_{{\cal G}'} e^{\beta \alpha N_V({\cal G}')} \ , \nonumber \\
\end{eqnarray}
where the sum over ${\cal G}'$ runs over all the graphs with $N_D +1$
dimers on the lattice.  Let ${\cal G}''({\cal G})$ be a graph formed
by adding an additional dimer to a graph ${\cal G}$. The additional
dimer is, of course, placed on one of the dimer vacancies AB in
the graph ${\cal G}$.
The number of vertices adjacent to A (excluding B) is $\sigma$, and
the number of vertices adjacent to B (excluding A) is also $\sigma$.
If $m$ of these $2\sigma$ vertices are vacant, then the number of
dimer vacancies in the graph ${\cal G}''({\cal G})$ is
$N_V({\cal G})-m-1$. If we look at all the different graphs
${\cal G}''({\cal G})$ which can be generated by adding a dimer to
a particular graph ${\cal G}$, $m$ assumes the values (1, .., 2$\sigma$)
with the respective probabilities $p(m)$ which we shall calculate.
Furthermore, since it is quite clear that the configurations of two 
small subsections of a large graph ${\cal G}$ which are distant from each other are
statistically independent, we assert that the probability distribution
$p(m)$ is the same for almost all graphs ${\cal G}$ (but $p(m)$
does depend on $N_D/N$ which has the same value for all the graphs ${\cal G}$).
Thus for almost all graphs ${\cal G}$ we have
\begin{eqnarray}
&& \sum_{{\cal G}''({\cal G})} e^{\beta \alpha N_V[{\cal G}''({\cal G})]}
\nonumber \\ &=& N_V({\cal G}) e^{\beta \alpha N_V({\cal G})}
\sum_{m=0}^{2\sigma} p(m) e^{-\beta \alpha (m+1)} \ .
\end{eqnarray}
If $\{ {\cal G}''({\cal G})\}$ is the set of all graphs which can be generated
by adding a dimer to a particular graph ${\cal G}$, and
$\sum_{\cal G} \{ {\cal G}''({\cal G})\}$ is the set which is the
union of all the sets $\{{\cal G}''({\cal G})\}$, then
$\sum_{\cal G} \{ {\cal G}'' ({\cal G})\}$ is identical to the set
of graphs $\{{\cal G}'\}$, except that each graph ${\cal G}'$ occurs
$N_D +1$  times in the union (since each graph ${\cal G}'$ has
$N_D +1$ "ancestors" ${\cal G}$ which can be obtained by
removing a single dimer). Thus we obtain
\begin{eqnarray}
&& (N_D+1) \sum_{{\cal G}'} e^{\beta \alpha N_V({\cal G}')} \nonumber \\
&=& \sum_{\cal G} \sum_{{\cal G}''({\cal G})} e^{\beta \alpha N_V[
{\cal G}''({\cal G})]} \nonumber \\
&=& \Biggl( \sum_{m=0}^{2 \sigma} p(m) e^{-\beta \alpha (m+1)} \Biggr)
\sum_{\cal G} N_V({\cal G}) e^{\beta \alpha N_V({\cal G})} \nonumber \\
&=& \Biggl( \sum_{m=0}^{2 \sigma} p(m) e^{-\beta \alpha (m+1)} \Biggr)
\overline N_V (N_D) \sum_{\cal G} e^{\beta \alpha N_V({\cal G})} \ , \nonumber \\
\end{eqnarray}
where
\begin{eqnarray}
\overline N_V(N_D) &=& \sum_{\cal G} N_V({\cal G})
e^{\beta \alpha N_V({\cal G})}/\sum_{\cal G} e^{\beta \alpha N_V({\cal G})} \ .
\end{eqnarray}
For a given value of $z$, the value of $N_D$  which maximizes $P(N_D)$
satisfies the condition $P(N_D+1)/P(N_D)=1$,
\begin{eqnarray}
1 = z e^{\beta \alpha (2q-1)} {\overline N_V(\overline N_D) \over
\overline N_D+1} \sum_{m=0}^{2 \sigma} p(m) e^{-\beta \alpha (m+1)} \ .
\label{INTEQ} \end{eqnarray}
(Note that when $\alpha=0$ this reduces to Eq. (\ref{NVNDEQ}) for
noninteracting dimers.)
The preceding statements are true on all lattices with $q$ nearest
neighbors. 

\begin{figure}
\psfig{figure=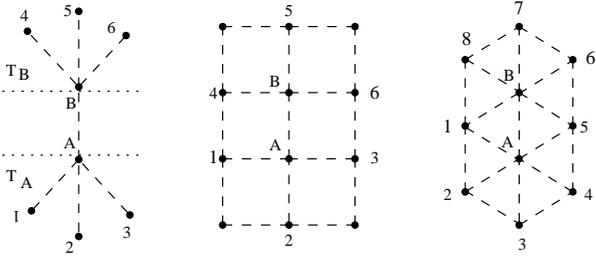}
\caption{Left: the bond AB and its environment in a Bethe lattice with $q=4$.
The subsystems T$_{\rm A}$ and T$_{\rm B}$ are the regions outside the
dotted lines.  We show the sites neighboring to A and B, which are
numbered 1 to 6 (1 to $2\sigma$).  There are no indirect paths connecting any
pairs of numbered sites which do not pass through A and/or B. Middle:
Environment of a bond AB on a square lattice where we show indirect
paths which connect adjacent numbered sites, in which case the strong
form of tree decoupling is not exact. Right: Environment of a bond
AB on a triangular lattice.  Here the decoupling is worse than for
a square lattice because a) there are nearest neighbor bonds
connecting numbered sites and b) some neighbors of A are simultaneously
neighbors of B.}
\label{TATB}
\end{figure}

We now obtain $N_V$ and $p(m)$ explicitly as functions of $N_D$, so that
Eq. (\ref{INTEQ}) becomes an explicit equation of state for dimers.
We do this using an approximation, which for reasons explained below
we call the ``strong form of tree decoupling,'' in which we replace the
$d$-dimensional lattice by a Bethe lattice. To obtain $p(m)$ we first
relate $p(m)$ to the quantity $p_{\rm con}$ which we define to be the
conditional probability that site $i$ is vacant when site $j$ is known
to be vacant, where sites $i$ and $j$ are nearest neighboring sites.
Recall that A and B are vacant neighboring sites.
We label the vertices adjacent to A (excluding B)
by the index $i$ ($i=1,..,\sigma$) and the vertices adjacent to B 
(excluding A) by the values $i=\sigma+1, \dots 2\sigma$, as shown in
Fig. \ref{TATB}. We define a
random variable $X_i$  which has the value 1 if the vertex $i$ is vacant
and the value 0 if a dimer is touching vertex $i$. 
On the Bethe lattice the $X_i$ ($i=1, 2, \dots 2 \sigma$)
are independent random variables. Note that $P(X_i=1)$ is the conditional
probability that a neighbor of B is vacant given that not only B but also
A is vacant.  On a Bethe lattice the condition that A is also vacant is irrelevant
in this context, in which case $P(X_i=1)$ is just $p_{\rm con}$, so that
\begin{eqnarray}
p(m)&=& P(X_1 + X_2 \dots +X_{2\sigma} = m) \nonumber \\
& =& {(2 \sigma)! \over m! (2\sigma-m)!} p_{\rm con}^m
(1-p_{\rm con})^{2 \sigma-m}
\label{POFMEQ} \end{eqnarray}
and
\begin{eqnarray}
&& \sum_{m=0}^{2 \sigma} p(m) e^{- \beta \alpha (m+1)}\nonumber \\
&=& e^{-\beta \alpha} [p_{\rm con}e^{-\beta \alpha} + 1
- p_{\rm con} ]^{2 \sigma} \ .
\label{SUMEQ} \end{eqnarray}
Thus, Eq. (\ref{INTEQ}) becomes
\begin{eqnarray}
1 = z {\overline N_V \over \overline N_D} [\epsilon+1-p_{\rm con}
\epsilon]^{2 \sigma} \ ,
\label{INTEQ2} \end{eqnarray}
where $\epsilon = e^{\beta \alpha}-1$ and we replaced
$\overline N_D+1$ by $\overline N_D$ in the thermodynamic limit.
Now we use Eq. (\ref{NVEQ}) which we write in the form
\begin{eqnarray}
\rho_V &=& (1-q \rho )p_{\rm con} \ .
\end{eqnarray}
Then Eq. (\ref{INTEQ2}) becomes
\begin{eqnarray}
1 = zp_{\rm con} {1-q \rho \over \rho} [\epsilon+1-p_{\rm con} \epsilon]^{2 \sigma} \ .
\label{INTEQ3} \end{eqnarray}
It remains to determine $p_{\rm con}$ in terms of $\rho$.

In order to calculate $p_{\rm con}$ (as a function of $\rho$), we first calculate
a simpler quantity $p'$, the conditional probability that B is vacant,
given that bond AB is unoccupied (i.e. not covered by a dimer). Note that
an unoccupied bond is not necessarily a dimer vacancy, since a dimer may
be touching one or both ends of the bond. We write
\begin{eqnarray}
p' &=& P({\rm B\ vacant} \ |\ {\rm bond\ AB\ unoccupied})\nonumber \\
&=& {P({\rm B\ vacant\ and\ bond\ AB\ unoccupied}) \over
P({\rm bond\ AB\ unoccupied})} \nonumber \\
&=&P({\rm B\ vacant})/P({\rm bond\ AB\ unoccupied}) \nonumber \\
&=& (1 - q \rho )/(1-\rho ) \ .
\label{PPEQ} \end{eqnarray}
On the other hand, we will calculate $p'$ in terms of the weights
(unnormalized probabilities) of certain graphs, yielding the
information which is needed to calculate $p_{\rm con}$.

We focus our attention on a bond AB and we continue to consider
the Bethe lattice.  For this discussion we divide the lattice into two parts
by cutting the bond AB, so that one part, T$_{\rm A}$, contains site
A and all sites accessible to site A without going through site B
and the other part, T$_{\rm B}$, is defined similarly, as shown in
Fig. \ref{TATB}. On the Bethe lattice, when the bond AB is unoccupied,
these two parts T$_{\rm A}$ and T$_{\rm B }$ are independent subsystems. 
In that case we see, from Eq. (\ref{PNDEQ}), that the "weight" of a configuration
of either T$_{\rm A}$ or T$_{\rm B}$ is obtained as the product of
(a) a factor $\zeta=z\exp [\beta \alpha (2q-1)]$
for every bond covered by a dimer\cite{FN2}
(the actual value of $\zeta$ will not appear in the final results)
and (b) a factor $e^{\beta \alpha}$ for every bond which is a
dimer vacancy.  We define
\begin{eqnarray}
w_{A,0} &=& {\rm sum\ of\ weights\ of\ all\ graphs\ on\ T}_{\rm A}
\ {\rm which} \nonumber \\ && \ \
\ {\rm contain\ no\ dimers\ touching\ A} \nonumber \\
w_{A,1} &=& {\rm sum\ of\ weights\ of\ all\ graphs\ on\ T}_{\rm A}
\ {\rm which} \nonumber \\ && \ \
 {\rm contain\ one\ dimer\ touching\ A} \nonumber \\
w_{B,0} &=& {\rm sum\ of\ weights\ of\ all\ graphs\ on\ T}_{\rm B}
\ {\rm which} \nonumber \\ && \ \
\ {\rm contain\ no\ dimers\ touching\ B} \nonumber \\
w_{B,1} &=& {\rm sum\ of\ weights\ of\ all\ graphs\ on\ T}_{\rm B}
\ {\rm which} \nonumber \\ && \ \
\ {\rm contain\ one\ dimer\ touching\ B} \ .
\end{eqnarray}
Note that we are only concerned with graphs in which the bond AB is
fixed (to be unoccupied), so that weights of graphs on the
subtrees $T_{\rm A}$ and $T_{\rm B}$ are well defined. Then
\begin{eqnarray}
\label{WEQ}
p' &=& {P({\rm B\ vacant}) \over P({\rm AB\ unoccupied})} \\
&=& {w_{A,0}w_{B,0} e^{\beta \alpha} + w_{A,1}w_{B,0} \over
w_{A,0}w_{B,0} e^{\beta \alpha} + w_{A,0}w_{B,1} +
w_{A,1}w_{B,1} + w_{A,1}w_{B,0}} \ . \nonumber
\end{eqnarray}
The first term in the numerator is the total weight of graphs in which
A and B are vacant (so that bond AB is a dimer vacancy), and the second
term is the total weight of graphs in which B is vacant and A is occupied
(so that bond AB is not a dimer vacancy). The terms in the denominator are
similarly interpreted. Since $P({\rm B\ vacant})=P({\rm A\ vacant})$
far from the boundary, we have $w_{B,0}w_{A,1}=w_{B,1}w_{A,0}$.
Thus $w_{A,1}/w_{A,0}=w_{B,1}/w_{B,0}$.

Dividing the numerator and denominator of Eq. (\ref{WEQ}) by
$w_{A,0}w_{B,0}$ we obtain the following relation between $p'$ and
$u\equiv w_{A,1}/w_{A,0}$:
\begin{eqnarray}
p' &=& {e^{\beta \alpha} + u \over e^{\beta \alpha} + u^2 + 2u} \ ,
\end{eqnarray}
so that
\begin{eqnarray}
{e^{\beta \alpha} + u \over e^{\beta \alpha} + u^2 + 2u} 
&=& {1- q \rho \over 1- \rho } \ .
\label{UEQ1} \end{eqnarray}
Now we relate $p_{\rm con}$ to $u$:
\begin{eqnarray}
p_{\rm con} &=& {P({\rm A\ and\ B\ vacant}) \over P({\rm A\ vacant})}
\nonumber \\ &=& {w_{A,0}w_{B,0} e^{\beta \alpha} \over
w_{A,0} w_{B,0} e^{\beta \alpha} + w_{A,0} w_{B,1}} \nonumber \\
&=& {\epsilon +1 \over \epsilon + 1 + u} \ .
\label{UEQ2} \end{eqnarray}
We now solve Eq. (\ref{UEQ2}) for $u$ and substitute the result into
Eq. (\ref{UEQ1}) to get
\begin{eqnarray}
\epsilon p_{\rm con} -(2 \epsilon+\lambda + 1) + (\epsilon+1)/p_{\rm con}
= 0 \ .
\label{POFYEQ} \end{eqnarray}
where $\lambda = \rho \sigma/(1-q \rho)$.  (Note that as $\alpha \rightarrow 0$,
$p_{\rm con}$ becomes equal to $p'$, as expected.) Thus we get
\begin{eqnarray}
{1-q \rho \over \rho} &=& {p_{\rm con} \sigma \over [\epsilon p_{\rm con}
- \epsilon - 1][p_{\rm con}-1]} \ ,
\end{eqnarray}
so that Eq. (\ref{INTEQ3}) can be written as
\begin{eqnarray}
1 = {\sigma zp_{\rm con}^2 \over 1-p_{\rm con}} 
[\epsilon+1-p_{\rm con} \epsilon]^{2 \sigma-1} \ ,
\label{PEQ} \end{eqnarray}
or
\begin{eqnarray}
- \ln(\sigma z) &=& G(\rho ) \ ,
\label{ZEQ} \end{eqnarray}
where
\begin{eqnarray}
G(\rho ) &=& 2 \ln p_{\rm con}(\rho ) - \ln[1-p_{\rm con}(\rho )] \nonumber \\ && \ 
+ (2\sigma -1) \ln [\epsilon +1 -\epsilon p_{\rm con}(\rho )] \ ,
\label{SELFC} \end{eqnarray}
where $p_{\rm con}(\rho )$ is determined by Eq. (\ref{POFYEQ}).\cite{1939}

\begin{figure}
\psfig{figure=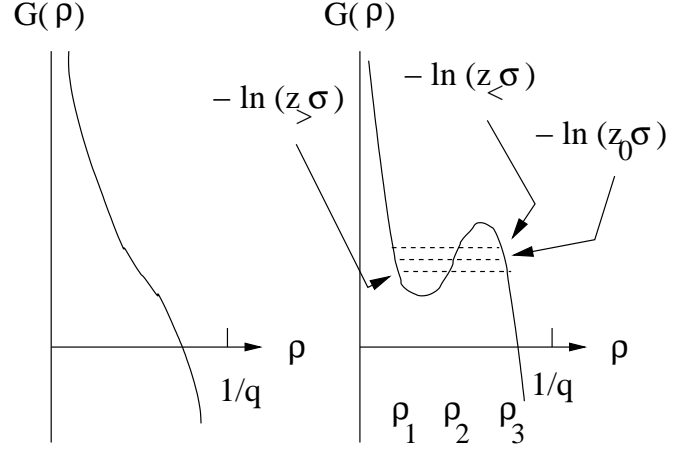}
\caption{Schematic graphs of $G(\rho )$ for $\epsilon< \epsilon_c$
(left) and for $\epsilon> \epsilon_c$ (right). In the right panel
we show the line at $- \ln [\sigma z_0(\epsilon)]$ at
which the net area between the
line and the curve of $G(\rho)$ is zero.  This is the ``equal area''
construction which gives $z_0(\epsilon)$.  We also show lines
for a value of $z$ ($z_<$) which is less than $z_0(\epsilon )$ and
a value of $z$ ($z_>$) which is larger than $z_0(\epsilon)$.
The values of $\rho_j$ are the values of $\rho$ at which
the line for $z_0(\epsilon)$ intersects $G(\rho)$.}
\label{FIG3} \end{figure}

The physical range of $\rho $ is $[0,1/q]$. As $\rho \rightarrow 0$, 
$p_{\rm con}\rightarrow 1$ and $G(\rho )\rightarrow  + \infty$.
As $\rho \rightarrow 1/q$, $p_{\rm con}\rightarrow 0$ and
$G(\rho ) - \infty$. One can show, either analytically or by making
computer plots, that if $\epsilon$ is less than a certain critical value
$\epsilon_c$ (which we shall calculate) then $G(\rho )$ is a monotone
decreasing function of $\rho$ in the interval $[0,1/q]$. In this case there
is a unique solution of Eq. (\ref{ZEQ}) for $\rho$ for each positive value
of the activity $z$. If $\epsilon > \epsilon_c$, then $dG/d\rho$ is 
negative for small $\rho$ and for $\rho$ near $1/q$, but there is an intermediate
region in which $dG/d\rho > 0$, as is illustrated in Fig. \ref{FIG3}.  In
this case there is a certain range of $z$ in which there are three values
of $\rho$ which satisfy Eq. (\ref{ZEQ}). As explained in Appendix
\ref{VDWAP}, for $z$ less than a certain value $z_0(\epsilon)$,
which is determined by an equal-area construction, the physical
solution (i.e. the most probable state) is
a homogeneous state whose density of dimers is given by the smallest
solution of Eq. (\ref{ZEQ}). For $z > z_0(\epsilon)$ the physical solution
is a homogeneous state whose density of dimers is given by the largest
solution of Eq. (\ref{ZEQ}).
For $z = z_0 (\epsilon)$ the physical solution is the coexistence of two 
homogeneous phases whose densities are the smallest and largest
solutions of Eq. (\ref{ZEQ}).

We now will calculate $\epsilon_c$ by determining the range of $\epsilon$ for
which the equation $G'(\rho )=0$ has two solutions for $\rho$ in $[0,1/q]$. We have
\begin{eqnarray}
G'(\rho ) &=&  \Biggl[ {2 \over p_{\rm con}} + {1 \over 1-p_{\rm con}}
- {\epsilon (2 \sigma -1) \over \epsilon + 1 - p_{\rm con} \epsilon }
\Biggr] {dp_{\rm con} \over d\rho } \ . \nonumber \\ 
\end{eqnarray}
By differentiating Eq. (\ref{POFYEQ}) with respect to $\rho$ one can show that
$dp_{\rm con}/d\rho <0$ for $p_{\rm con}$ in $[0,1]$ and $\rho $ in $[0,1/q]$.  Accordingly,
setting $G'(\rho )=0$ leads to the quadratic equation
\begin{eqnarray}
2 \sigma \epsilon p_{\rm con}^2 - p_{\rm con}(1 + 2 \epsilon
+ 2 \sigma \epsilon) + 2 \epsilon + 2 = 0 \ .
\label{QUADEQ} \end{eqnarray}
This equation has two real roots for either $\epsilon > \epsilon_+$ or
$\epsilon < \epsilon_-$ and no real roots otherwise, where $\epsilon_\pm$
satisfies
\begin{eqnarray}
( 1 + 2 \epsilon + 2 \sigma \epsilon )^2
= 16 \sigma \epsilon  (\epsilon+1) \ ,
\end{eqnarray}
which gives
\begin{eqnarray}
\epsilon_\pm &=& {3 \sigma-1 \pm  2 \sqrt{\sigma(2\sigma-1)} 
\over 2 (\sigma-1)^2 } \ . 
\label{EPSCEQ} \end{eqnarray}
Since the two values of $p_{\rm con}$ corresponding to $\epsilon_-$ are
both greater than 1, we see that two relevant solutions for $G'(\rho )=0$
occur only for $\epsilon> \epsilon_c \equiv \epsilon_+$.
From the value of $\epsilon_c$ we obtain the transition temperature $T_c$
for the liquid-gas transition for interacting dimers from
\begin{eqnarray}
\epsilon_c = e^{\alpha /(kT_c)} -1 \ .
\label{ECEQ} \end{eqnarray}

We close this section by discussing the accuracy of
the ``strong form of tree decoupling`` in which we
approximate a $d$-dimensional lattice by a Bethe lattice of the
same coordination number.  Recall that in the tree decoupling
introduced above Eq. (\ref{NV1EQ}) it was only assumed that we could neglect
indirect paths which connect nearest neighboring sites A and B.
A comparable approximation here is to assume that the conditional
probability that a neighbor of site A is vacant given that both
sites A and B are vacant is the same as the conditional probability
when only site A is vacant.  If this were the only approximation,
then the present treatment of the interacting dimer system would
be expected to be quite accurate - as is the tree decoupling for
noninteracting dimers.\cite{CC1}  However, here we also started from
the much stronger assumption that the random variables $X_i$ introduced above
Eq. (\ref{POFMEQ}) are independent of one another.  On the Bethe lattice
this is obviously true, as can be seen from Fig. \ref{TATB}.  However,
on a real lattice, this assumption neglects the many next-nearest
neighbor connections between the $X_i$ sites.  This same neglect reappears
if one tries to apply the subsequent calculation of $p'$ to a
$d$ dimensional lattice.  Accordingly it would not be surprising if
the Bethe approximation for interacting dimers had an accuracy
similar to that of mean field theory for the Ising model in the
same spatial dimensionality.  In two dimensions, this would imply that
the Bethe lattice value of $\epsilon_c$ may differ from the
exact value for a $d$ dimensional lattice by about 30\%. 
 
\subsection{Solution by Construction of an Effective Hamiltonian} 

In this section we generalize the effective Hamiltonian for noninteracting
dimers to include dimer-dimer interactions.  We continue to mark sites with
operators $s_i$ which obey the trace rules of Eq. (\ref{RULES}).
But in addition we need to have operators which keep track of interactions.
So at each site $i$ we introduce operators $t_i$ which have zero trace
unless accompanied by an $s_i$ operator.  These operators commute with one
another and obey
\begin{eqnarray}
{\rm Tr}_j s_j^p t_j^r =\delta_{p,0} \delta_{r,0} + \delta_{p,1} 
\end{eqnarray}
for $p$ and $r$ each assuming the values $0, 1, \dots q$.  Now we set
\begin{eqnarray}
e^{- \beta \cal H}
&=& \prod_{\langle ij \rangle} \Biggl[ \left( 1 + x s_i s_j \right)
\left( 1 + \delta t_i t_j \right) \Biggr] \ .
\end{eqnarray}
For a configuration of $N_D$ dimer bonds  and $N_I$ interacting bonds
this Hamiltonian gives a contribution to the partition function of
\begin{eqnarray}
[x(1+\delta)]^{N_D} [1+\delta]^{N_I} \ ,
\end{eqnarray}
To get the desired partition function we thus set $\delta=e^{\alpha \beta}-1 = \epsilon$
and $x=ze^{-\beta \alpha}$.

We now obtain the exact solution for the Bethe lattice.
We could introduce replicas to obtain an expansion for
the free energy rather than for the partition function.  However, in the
interest of simplicity we work with the partition function to obtain
an exact solution for the Bethe lattice.  Accordingly Eq. (\ref{NONLIN})
in this case assumes the form
\begin{eqnarray}
g_i &=& {{\rm Tr}_j \Biggl[ \left( 1 + ze^{-\beta \alpha} s_i s_j \right)
\left( 1 + \epsilon t_it_j \right) g_j^\sigma \Biggr] \over {\rm Tr}_j g_j^q } \ .
\label{INTNON} \end{eqnarray}
The solution to this equation is of the form
\begin{eqnarray}
g_i &=& A + B s_i + C t_i + D s_i t_i \ .
\end{eqnarray}
To satisfy Eq. (\ref{INTNON}) the constants must obey
\begin{eqnarray}
A &=& {\rm Tr}_j g_j^\sigma / {\rm Tr}_j g_j^q \nonumber \\ 
B &=& z e^{-\beta \alpha} {\rm Tr}_j s_j g_j^\sigma / {\rm Tr}_j g_j^q \nonumber \\ 
C &=& \epsilon {\rm Tr}_j t_j g_j^\sigma / {\rm Tr}_j g_j^q \nonumber \\ 
D &=& z \epsilon e^{-\beta \alpha}
{\rm Tr}_j s_jt_j g_j^\sigma / {\rm Tr}_j g_j^q \ .
\end{eqnarray}
These equations may be written in terms of $A$, $r_B \equiv B/A$,
$r_C \equiv C/A$ and $r_D=D/A$, as
\begin{eqnarray}
A^2 &=& {1 + \sigma (r_B + r_D)(1 + r_C)^{\sigma-1} \over
1 + q(r_B + r_D)(1 + r_C)^\sigma } \nonumber \\
r_B &=& {ze^{-\beta \alpha} (1 + r_C)^\sigma \over
1 + \sigma (r_B + r_D)(1 + r_C)^{\sigma-1} } \nonumber \\
r_C &=& {\epsilon \sigma (r_B+r_D) (1 + r_C)^{\sigma-1} \over
1 + \sigma (r_B + r_D)(1 + r_C)^{\sigma-1} } \nonumber \\
r_D &=& {z \epsilon e^{-\beta \alpha} (1 + r_C)^\sigma \over
1 + \sigma (r_B + r_D)(1 + r_C)^{\sigma-1} } =  \epsilon r_B \ .
\end{eqnarray}
From the equation for $r_C$ we obtain
\begin{eqnarray}
r_B + r_D &=& {r_C \over \sigma (1+ r_C)^{\sigma-1} (\epsilon - r_C)} \ .
\end{eqnarray}
Using this in conjunction with the equation for $r_B$ we obtain an
equation which determines $r_C$:
\begin{eqnarray}
{r_C/(1 + \epsilon) \over \sigma (\epsilon- r_C)(1+r_C)^{\sigma-1}}
&=& {z (1+r_C)^\sigma \over (1+\epsilon)[1 + r_C/(\epsilon- r_C)]}
\nonumber \\
\end{eqnarray}
which can be put into the form
\begin{eqnarray}
-\ln (\sigma z) &=& 2 \ln (\epsilon-r_C) - \ln (\epsilon r_C)
\nonumber \\ && \ + (2\sigma-1) \ln (1+ r_C) \ .
\end{eqnarray}
This is identical to (\ref{ZEQ}) when we make the identification
$r_C/\epsilon= p_{\rm con}$.  (It has to be admitted that this
physical interpretation of $r_C$ is not obvious if one only has the
effective Hamiltonian.)

In Appendix \ref{MFAP} we develop mean field theory by a suitable
decoupling of the effective Hamiltonian and, as expected, we
 obtain results qualitatively similar to those for the Bethe lattice.

\section{QUENCHED RANDOMNESS}

\subsection{Dimers on Percolation Clusters at Infinite Fugacity}

Here we consider the statistics of dimers on a quenched random lattice
in which sites can be either  ``X'' sites with probability $p$
or ``Y'' sites with probability $1-p$.  For a given configuration ${\cal C}$
(i. e. for a given distribution of X and Y sites), the grand canonical
partition function, $Z({\cal C};\{z\})$ for dimer coverings is calculated as
\begin{eqnarray}
Z({\cal C};\{z\}) &=& \sum_{\cal A} z_{XX}^{n_{XX}({\cal A})}
z_{XY}^{n_{XY}({\cal A})} z_{YY}^{n_{YY}({\cal A})} \ , 
\end{eqnarray}
where the sum is over all arrangements ${\cal A}$ of 0, 1, 2, $\dots$
hard-core dimers, $n_{AB}({\cal A})$ is the number of dimers covering
an A site and a B site in the arrangement ${\cal A}$
(where A and B each assume the values X and Y), 
and $z_{AB}$ is the activity of an AB dimer.
Then the quenched average free energy $F$ is calculated as
\begin{eqnarray}
F(p;\{z\}) = \sum_{\cal C} P({\cal C};p) \ln Z({\cal C};\{z\})\ ,
\end{eqnarray}
with $P({\cal C};p) = p^{n_x({\cal C})} (1-p)^{n_y({\cal C})}$, where
$n_x({\cal C})$ ($n_y({\cal C})$) is the number of X (Y) sites in the
configuration ${\cal C}$.  From this quenched free energy one then
obtains the average number of AB dimers as
\begin{eqnarray}
N_{AB}(p;\{z\}) &=& z_{AB} {\partial F(p;\{z\}) \over \partial z_{AB}} \ .
\end{eqnarray}

An exact result for the dimer density on a Bethe lattice does not seem
easy to obtain.  However we now give an exact solution for the dimer density
on a Bethe lattice in the limit $z_{XX} \rightarrow \infty$ and
$z_{XY}=z_{YY} \rightarrow 0$.  Thus we consider $M_D(p)$, the average
of the maximal number of dimers which can be placed on percolation
clusters of X sites. Even in this limit the result is not
trivial because the constraint that dimers do not overlap plays a
variable role depending on  the compactness of the cluster.

\begin{figure}
\centerline{\psfig{figure=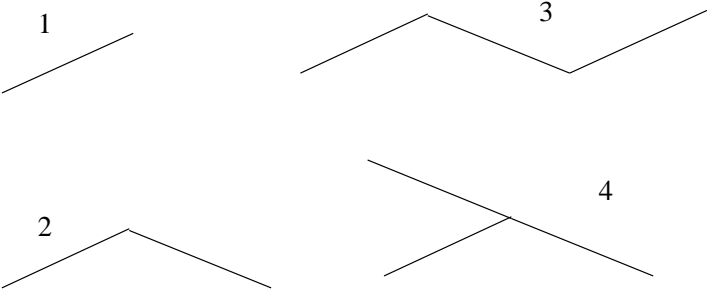}}
\caption{Small clusters for which we give the cumulant value of $M$.}
\label{FIG2}
\end{figure}

To obtain the exact solution we will explicitly evaluate the expansion of
$M_D(p)$ in powers of $p$.  For this purpose we temporarily consider
$M_D$ as a function of the set of $p_i$'s, where $p_i$ is the probability
that site $i$ is present (i. e. is an X site).  Then we write
\begin{eqnarray}
&& M_D(\{p_i\}) = \sum_i M_D(0, ... 0,p_i,0, ... 0) \nonumber \\
&& \ \ \ \ \  + \sum_{i<j} \Biggl[ M_D(0,...0,p_i,0,...,0,p_j,0,...0) \nonumber \\
&& \ \ \ \ \  - M_D(0,...0,p_i,0,...0) -M_D(0,...0,p_j,0,...0)
\Biggr]\nonumber \\ && \ \ \ \ \ + \dots \ .
\end{eqnarray}
We write this as
\begin{eqnarray}
M_D(\{p_i\}) &=& \sum_i M_D(p_i)_c + \sum_{i<j} M_D(p_i,p_j)_c \nonumber \\
&& \ \ + \sum_{i<j<k} M_D(p_i,p_j, p_k)_c \dots \ , \nonumber \\
\label{127} \end{eqnarray}
where we only indicate as arguments those $p_i$'s which are nonzero and 
we introduce the cumulants via
\begin{eqnarray}
M_D(p_i)_c &=& M_D(p_i) \nonumber \\
M_D(p_i,p_j)_c &=& M_D(p_i,p_j) - M_D(p_i) - M_D(p_j) \ , \nonumber \\
\end{eqnarray}
etc., as in Eq. (47).
Note that $M_D(p_i)=0$ because a dimer requires two sites being present.
Similarly $M_D(p_i,p_j)\not= 0$ only if sites $i$ and $j$ are nearest neighbors.
On can show that the cumulant vanishes for a disconnected diagram.
The general term (evaluated for $p_i=p$) is
\begin{eqnarray}
M_D(\Gamma)_c &=&
\sum_{\gamma \in \Gamma} M_D(\gamma) (-1)^{N_\Gamma - N_\gamma}
\nonumber \\ & =& {\rm The\ terms\ of\ order\ } p^{N_\Gamma}
\ {\rm in}\ M_D(\Gamma)
\nonumber \\ &=& p^{N_\Gamma} \sum_{x_i = \pm 1}
M_D( x_1, x_2 ,\dots x_{N_\Gamma}) \prod_{j=1}^{N_\Gamma} x_j \ . \nonumber \\ 
\label{CUMEQ} \end{eqnarray}
In the first line $\Gamma$ denotes a set of $p_i$'s, $\gamma$ is
a subset of $\Gamma$ (with $\gamma = \Gamma$ allowed), and $N_\Gamma$
is the number of $p_i$'s in the set $\Gamma$.  In the last line
$x_i=1$ means that the site $i$ is included in the set and
$x_i=-1$ means that the site $i$ is not included.  (These definitions
follow from the fact that to get a term of order $p^{N_\Gamma}$,
we take a factor of $p$ if the site is included and 
$(1-p)\rightarrow (-p)$ if the site is not included.)
For example, for the sets of sites shown in Fig. \ref{FIG2}
we find that 

\begin{eqnarray}
M_D(\Gamma_1)_c &=& p^2 \ , \nonumber \\
M_D(\Gamma_2)_c &=& p^3 [ 1 -2(1)] =-p^3  \ , \nonumber \\
M_D(\Gamma_3)_c &=& p^4 [ 2 - 4(1) + 3(1)] = p^4 \ , \nonumber \\
M_D(\Gamma_4)_c &=& p^4 [ 1 - 3(1) +3(1)]=p^4 \ .
\end{eqnarray}
For $\Gamma_2$ there is one term with all the $x$'s equal to $+1$ and
two nonzero terms with one $x_i$ equal $-1$.  For $\Gamma_3$ there is
one term with all the $x$'s equal to $+1$, four nonzero terms with
one $x_i$ equal $-1$, and three nonzero terms with two $x_i$'s equal to
$-1$.  For $\Gamma_4$ there is one term with all the $x$'s equal to $+1$,
three nonzero terms with one $x_i=-1$, and three nonzero terms with
two $x_i$'s equal to $-1$.  These results suggest that for any connected
cluster of sites $\Gamma$ on the Bethe lattice one has
\begin{eqnarray}
M_D(\Gamma)_c &=& (-p)^{N_\Gamma} \ .
\label{PROOFEQ} \end{eqnarray}
Note that in contrast to the cumulant, the bare value $M_D(\Gamma)$
is not simply a function of $N_\Gamma$.

The proof of Eq. (\ref{PROOFEQ})
is by induction on $N_\Gamma$.  We have explicitly shown this
result to be true for $N_\Gamma$ equal to 2, 3, and 4.  We now show that
if Eq. (\ref{PROOFEQ}) is assumed to hold for $N_\Gamma \leq N-1$, then
it holds for $N_\Gamma=N$ (assuming $N>2$).  When this is proved, the
general result is established.

Consider a diagram with $N$ sites and label the sites so that
the $N$th site is an ``end'', that is, it is connected to only one other
site in the diagram and this other site is labeled $N-1$, as shown in
Fig. 7.  For a Bethe lattice (in contrast to the case of hypercubic
lattices) this construction is possible because all diagrams have at
least one free end. 

\vspace{0.2 in}
\begin{figure}
\centerline{\psfig{figure=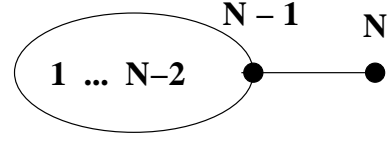}}
\caption{Cluster of $N$ sites in which site $N$ is a free end.}
\label{PR1.FIG} \end{figure}

\vspace{0.2 in} \noindent
We will use Eq. (\ref{CUMEQ}) for the cumulant, so that
\begin{eqnarray}
M_D(\Gamma_N)_c = p^N \sum_{x_1 , x_2 \dots x_N=\pm 1}
x_1 x_2 \dots x_N F(x_1, x_2, \dots x_N) \ , \nonumber \\
\end{eqnarray}
which we write as
\begin{eqnarray}
M_D(\Gamma_N)_c &=& p^N \sum_{x_1 , x_2 \dots x_{N-2}} x_1 x_2 \dots x_{N-2}
\nonumber \\ && \ \times [A-B+C-D] \ ,
\end{eqnarray}
where the algebraic signs reflect the values of the quantity $(x_{N-1}x_N)$
with
\begin{eqnarray}
A &=& M_D(x_1 , x_2 \dots x_{N-2}, 1, 1) \nonumber \\
B &=& M_D(x_1 , x_2 \dots x_{N-2}, 1, -1) \nonumber \\
C &=& M_D(x_1 , x_2 \dots x_{N-2}, -1, -1) \nonumber \\
D &=& M_D(x_1 , x_2 \dots x_{N-2}, -1, 1) \ .
\end{eqnarray}
In terms $B$, $C$, and $D$, the maximal covering by dimers does {\it not}
include the bond $N-1$ to $N$, because for this bond to be included obviously
both sites $N-1$ and $N$ must be occupied (so that $x_{N-1}=x_N=1$). Consider 
\begin{eqnarray}
Y(\Gamma_N) = p^N \sum_{x_1 , x_2 \dots x_{N-2}} x_1 x_2 \dots x_{N-2} [-B+C] \ .
\end{eqnarray}
This is almost the contribution to the cumulant when the site $N$ is not
included either in the summations or in the covering. We say "almost"
because the sum includes the factor $-px_N$ which ought to be taken out
if we want to identify this with the $N-1$ site problem.  We have
\begin{eqnarray}
p^N && \sum_{x_1 , x_2 \dots x_{N-2}} x_1 x_2 \dots x_{N-2} [-B+C] 
\nonumber \\
&=& (-p)p^{N-1}  \sum_{x_1 , x_2 \dots x_{N-2}} x_1 x_2 \dots x_{N-2} [B-C] 
\nonumber \\ &=& (-p) M_D(\Gamma_{N-1}^\prime )_c \ ,
\end{eqnarray}
where $\Gamma_{N-1}^\prime$ is the diagram obtained from $\Gamma_N$
by omitting the $N$th site.

Now consider $A$.  Suppose the maximal covering
does not actually include the bond $N-1$ to $N$.  Then clearly, if this covering
is to be maximal it must include a bond from some site to $N-1$.  But now
we may keep the number of dimers maximal by moving this bond which includes the
site $N-1$ to cover the bond $N-1$ to $N$.  So, without loss of generality,
in the term $A$ the maximal covering can be chosen so
as to include the bond $N-1$ to $N$.  Thus for $N>1$,
\begin{eqnarray}
p^N && \sum_{x_1 , x_2 \dots x_{N-2}} x_1 x_2 \dots x_{N-2} A
\nonumber \\ &=& p^N \sum_{x_1 , x_2 \dots x_{N-2}}
x_1 x_2 \dots x_{N-2} \nonumber \\ && \ \times [M_D(x_1, x_2, \dots x_{N-2}) + 1] \ . 
\end{eqnarray}
Now if $N>2$ the term with the 1 will vanish when summed over
$x_1$.  (For $N=2$ there are no summations left and this is a special case.)
Then we see that for $N>2$
\begin{eqnarray}
p^N \sum_{x_1 , x_2 \dots x_{N-2}} x_1 x_2 \dots x_{N-2} A
&=& p^2 M_D(\Gamma_{N-2}^{\prime \prime} )_c \ , \nonumber \\
\end{eqnarray}
where $\Gamma_{N-2}^{\prime \prime}$ is obtained from $\Gamma_N$ by 
deleting sites $N$ and $N-1$.  The same reasoning can be applied to term D, but
in this case there is no term with 1.  So
\begin{eqnarray}
p^N \sum_{\{x_i\}} x_1 x_2 \dots x_{N-2} (-D)
&=& - p^2 M_D(\Gamma_{N-2}^{\prime \prime})_c \ , \nonumber \\
\end{eqnarray}
Combining all our results we have (for $N>2$)
\begin{eqnarray}
M_D(\Gamma_N)_c &=&
p^N \sum_{x_1 , x_2 \dots x_{N-2}} x_1 x_2 \dots x_n [A-B+C-D]
\nonumber \\ &=& (-p) M_D(\Gamma_{N-1}^\prime )_c \ .
\label{IND} \end{eqnarray}
This completes the proof by induction.

For a general lattice we may make the tree approximation in which the sum
is carried over all diagrams with no loops.  In that case the result is
that the average dimer density is given by
\begin{eqnarray}
M(p) &\equiv& M_D(p)/N_0 = N_0^{-1} \sum_{\cal C} (-p)^{N_{\cal C}} \ ,
\end{eqnarray}
where $N_0$ is the total number of sites, the sum is over all loopless
clusters ${\cal C}$ of 2 or more sites, and $N_{\cal C}$ is the
number of sites
in that cluster.  Thus for the Bethe lattice Eq. (\ref{127}) yields
\begin{eqnarray}
M(p) &=& \sum_{n>1} (-p)^n W(n) \ ,
\end{eqnarray}
where $W(n)$ is the number (per site) of clusters of $n$ sites on a
Bethe lattice.
(To avoid edge effects it may be more precise to say that $nW(n)$ is the
number of clusters of $n$ sites, one site of which is the central site
in an arbitrarily large Cayley tree.  This definition ensures that
this result applies to hypercubic lattices in the asymptotic limit
of large dimensionality.) We use the result of Fisher and Essam:\cite{JMP2}
\begin{eqnarray}
W(n+1)&=& {(\sigma+1) (\sigma n + \sigma)! \over (n+1)! (\sigma n + \sigma -n +1)!} \ .
\end{eqnarray}
Therefore the exact result can be written as
\begin{eqnarray}
M(p) & \equiv & \sum_{n > 1}^\infty (-p)^n {(\sigma+1) (\sigma n)! \over
n! (\sigma n -n +2)! } \ .
\label{EXACTEQ} \end{eqnarray}
This result indicates that the singularity in $M(p)$ is at $p=-p_c$, where
$p_c=(\sigma-1)^{\sigma-1}/\sigma^\sigma$ is the critical concentration for
branched polymers on the Bethe lattice.\cite{ABH1}  Curiously then,
dimer statistics on percolation  clusters is related to a somewhat artificial
model of localization (which is also related
to branched polymer statistics in the same way.\cite{ABH1})

For $\sigma=1$ (a linear chain) this gives
\begin{eqnarray}
M(p) = {p^2 \over 1 + p} \ .
\label{1DEQ} \end{eqnarray}
This result could be obtained far more simply by noting that the
expected number of clusters per site of length $n$ sites is $p^n(1-p)^2$
and for a cluster of length $2n$ or $2n+1$ the maximal number of dimers
is $n$, so that
\begin{eqnarray}
M(p) = \sum_{n=1}^\infty \left( p^{2n} + p^{2n+1} \right) (1-p)^2 n \ ,
\end{eqnarray}
which reproduces the result of Eq. (\ref{1DEQ}).

\subsection{Alternate Expression}

We now give a closed form expression for $M(p)$.  To do that we construct an
expression for $d^2M/dp^2$.  In this quantity the binomial coefficient can
be expressed as a contour integral:
\begin{eqnarray}
{d^2 M \over dp^2} &=& (\sigma+1) \sum_{n=0}^\infty
(-p)^n \int {dz \over 2 \pi i} { (1+z)^{\sigma n} \over z^{n-1} } \ ,
\end{eqnarray}
where the contour surrounds the origin.
The sum over $n$ is a geometric
series which can be summed.  After some manipulations we found that
\begin{eqnarray}
M(p) &=& p + z_0 - {\sigma-1 \over 2} z_0^2 \ ,
\end{eqnarray}
where $z_0$ is the root of the algebraic equation
\begin{eqnarray}
(1+z_0)^\sigma=-z_0/p \ ,
\end{eqnarray}
which is proportional to $p$ for small $p$.
One can check that this gives the correct result for $\sigma=1$.
For $\sigma=2$ it gives
\begin{eqnarray}
M(p) &=& {1 \over 4p^2} \Biggl[ -1 -6p -6p^2 +4p^3
+(1+4p)^{3/2} \Biggr] \ . \nonumber \\
\end{eqnarray}
This result shows that there is no singularity on the positive $p$ axis.

\section{RANDOM DEPOSITION}
\subsection{Introduction}

Here we consider a special case of a model of catalysis\cite{DLA}
in which hard-core dimers are randomly deposited on bonds
and the hard-core constraint does not allow two dimers to intersect
the same site.  Deposition on all allowable bonds is equiprobable.
A quantity of interest is the final concentration when no further
dimers can be deposited.  We have not been able to construct such a
solution on a Bethe lattice.  However, here we give an exact
solution of this model in one dimension.

To characterize this process it is essential to consider deposition
on a line of bonds of finite length in  one dimension.  In order
to consider deposition recursively, we therefore introduce
the function $F(i,j)$, (with $i<j$) which is defined to be the
average number of dimers which will ultimately be deposited in the
interval between two preexisting dimers, one on bond $i$ and the
other on bond $j$. Obviously $F(i,j)=F(i-n,j-n)$ and as shorthand
we set $F(1,j) \equiv F(j)$.

Let us see what this function is for small argument.  It
is clear that $F(1,3)=F(1,4)=0$ because a new dimer can not
be deposited on a bond neighboring an occupied bond, since
neighboring bonds share a site which can not be occupied by two
hard-core dimers.  Next consider $F(1,5)\equiv F(5)$.
If we start with dimers on bonds \#1 and \#5,
then an additional dimer can only be deposited on bond \#3, so 
\begin{eqnarray}
F(1,5)=F(5)= 1 \ 
\end{eqnarray}
is the average number of dimers which will be deposited between dimers at sites
\#1 and \#5.  Similar considerations indicate that $F(6)=1$ and $F(7)=5/3$. 

Next consider $F(1,8)$.  In the first step an additional dimer will be placed
on bonds \#3, \#4, \#5, or \#6, each with probability 1/4.
If it is placed on bond \#3 (or equivalently on bond \#6),
then we have added one dimer and will be able to add (on average)
$F(1,3)+F(3,8)=0+1=1$ further dimer in later step(s).  Therefore in each case
these two processes lead to the deposition of two dimers.  So the combined
contribution to $F(1,8)$ from these two cases, each occurring with probability
(1/4) is $\delta F(1,8)=1$.  Similarly, if the first new dimer is placed
on bond \#4 (or equivalently on bond
\#5), then we have added one dimer and will be able to add (on average)
$F(1,4)+F(4,8)=0+1=1$ further dimer.  As before, the contribution to
$F(1,8)$ from these two cases, each occurring with probability
(1/4) is $\delta F(1,8)=1$.  Therefore we see that $F(8)=2$.

\subsection{Recursion Relation}

Now imagine starting with dimers on bonds \#1 and $N$, with $N>4$.
The first added dimer can go on bond \#3, \#4, .... $N-3$, $N-2$,
each with probability $1/(N-4)$. So, if we include this added dimer
we have
\begin{eqnarray}
F(1,N) &=& 1 + {1 \over N-4} \sum_{j=3}^{j=N-2} [ F(1,j) + F(j,N)]
\nonumber \\ &=& 1 + {2 \over N-4}  \sum_{j=3}^{j=N-2} F(1,j)
\nonumber \\ &=& 1+ {2 \over N-4}  \sum_{j=3}^{j=N-2} F(j) \ .
\end{eqnarray}
Also
\begin{eqnarray}
F(N-1) &=& 1 + {2 \over N-5}  \sum_{j=3}^{j=N-3} F(j) \ .
\end{eqnarray}
Using these we see that
\begin{eqnarray}
(N-4) F(N) &=& (N-5)F(N-1) + 1 \nonumber \\ && \ + 2 F(N-2) \ .
\label{EQ2}\end{eqnarray}
Now we form the generating function
\begin{eqnarray}
\overline F(x) &=& \sum_{N=5}^\infty F(N) x^N \ .
\end{eqnarray}
Now multiply Eq. (\ref{EQ2}) by $x^N$ and sum from $N=5$ to $N=\infty$.
Keeping in mind that $F(N)$ vanishes for $N<5$ we get 
\begin{eqnarray}
x \overline F_x - 4 \overline F &=&
x^2 \overline F_x - 4x \overline F + {x^5 \over 1-x} +2x^2 \overline F \ ,
\end{eqnarray}
where $\overline F_x \equiv d\overline F(x) / dx$.   From this we find that
\begin{eqnarray}
\overline F(x) &=& {x^4 \over (1-x)^2} \Biggl(
{1 - e^{-2x} \over 2} \Biggr) \ .
\end{eqnarray}
To see what this implies about $F(N)$ for large $N$ we write
\begin{eqnarray}
\overline F(x) &=& {H(x) \over (1-x)^2} \nonumber \\
&=& {H(1) \over (1-x)^2 } - {H_1(1) \over 1-x} \nonumber \\ && 
\ \ + \sum_{n=2}^\infty {H_n(1) \over n!} (x-1)^{n-2} \ ,
\end{eqnarray}
where $H_n(1) \equiv d^nH(x)/dx^n|_{x=1}$.  From this we see that
\begin{eqnarray}
F(N) \sim H(1) N = ( 1 - \delta ) (N/2) \ ,
\end{eqnarray}
where
\begin{eqnarray}
\delta = e^{-2} \approx 0.1354 
\end{eqnarray}
is the fraction of sites which remain vacant in the jamming limit
(after deposition is completed).

\section{SUMMARY}

In this paper we have presented two approaches, which we call
geometric and algebraic, for the analysis of dimer statistics.
Both approaches yield exact results when the lattice does not have
loops, as for a Bethe lattice.  The charm of the geometrical method is that
it starts from the most basic statement, namely that in a grand canonical
ensemble, the activity $z$ is precisely the ratio of the statistical weights
of the $N$-particle system to that of the $N+1$-particle system. The virtue of
the algebraic approach is that once the Hamiltonian is constructed using operators
whose trace rules incorporate kinematic restrictions, the standard procedures
of statistical mechanics can be applied, for instance to obtain series
expansions for finite dimensional lattices.  (This approach was
used\cite{YS} to give a field theoretic analysis of the monomer-dimer
problem.) Here we also show that the use of replicas
(which normally are invoked to implement quenched averages of a random
Hamiltonian) can be useful in converting a series for the partition function
into one for its logarithm, the free energy.  Exact solutions for generalizations
in which a) the lattice is anisotropic or b) dimer-dimer interactions are
included were also developed using both approaches.  We give an exact
solution for dimer statistics on a Bethe lattice in a simple quenched
random potential in which dimers are placed on percolation clusters.
Finally, we developed an exact solution for the fraction of sites which remain
vacant after random disposition is completed in a one dimensional system.

\vspace{0.2 in} \noindent {\bf ACKNOWLEDGEMENTS}
We thank J. F. Nagle and M. E. Fisher for helpful correspondence.

\begin{appendix}

\section{Analysis of the Phase Transition}
\label{VDWAP}

The phase transition in the presence of dimer-dimer interactions is entirely
similar to that in the van der Waals gas.

The right-hand side of Eq. (\ref{PEQ}) gives an explicit formula for 
$zP(N_D+1)/P(N_D)$ when Eq. (\ref{POFYEQ}) is used.\cite{FN2}  Thus we have
\begin{eqnarray}
\ln [P(N_D+1)/P(N_D)] &=& G(\rho ) + \ln(\sigma z) \ .
\end{eqnarray}
Figure \ref{FIG3} shows schematic graphs of $G(\rho )$ versus $\rho$ for $\epsilon$
slightly less than and slightly greater than $\epsilon_c$.  For $\sigma=2$
we find $\epsilon_c \approx 4.95$ and the value of $\rho$ at the critical
point is $\rho_c \approx 0.159$, $G(\rho_c)=3.226$, and $z_c \approx 0.0199$.

If $\epsilon > \epsilon_c$ and $z$ is close to $z_c$, a horizontal line
(dashed in Fig. \ref{FIG3}) at height $\ln (1/\sigma z)$ will intersect the graph of
$G(\rho )$ three times, each of the intersections corresponding to a value of
$N_D$ for which $P(N_D+1)/P(N_D)=1$.  If we denote the intersections
$y_1<y_2<y_3$ and the corresponding density of dimers $\rho_1$, $\rho_2$, and
$\rho_3$, then we have
\begin{eqnarray}
\ln [P(\rho_3)/P(\rho_1)] = {Nq \over 2} \int_{\rho_1}^{\rho_3}
\left[ G(\rho) - \ln (1/\sigma z) \right] d\rho  \ .
\end{eqnarray}

If $z_0(\epsilon)$ is the value of $z$ such that the area of the
loop below the dashed line at height $\ln (1/\sigma z)$ is equal to
the area of the loop above that line, then $P(\rho_3)/P(\rho_1)=1$.  If $z$
is slightly less than $z_0$, then the negative area exceeds the positive
area and $P(\rho_3)/P(\rho_1)<1$, in which case is the stable phase has
$\rho=\rho_1$.
Note that $\rho_2$ is always less probable than $\rho_1$.  Similarly, if
$z$ is slightly greater than $z_0$ $z=z_<$, then the stable (most probable)
phase corresponds to $\rho=\rho_3$.  Note that because of the prefactor $Nq/2$
before the integral, ``most probably'' means ``overwhelmingly most probable.''

If we place a density of dimers on the lattice which is intermediate between
$\rho_1(z_0)$ and $\rho_3(z_0)$, the lattice will separate into regions with
coverage at densities $\rho_1(z_0)$ and $\rho_3(z_0)$, the size of the
two regions being determined by the requirement that the total number of
dimers is that specified.  
 
If $\epsilon$ is slightly greater than $\epsilon_c$, then the temperature
$T$ is slightly less than $T_c$ and the definition of $\epsilon$ implies
that $(T_c-T) \propto (\epsilon- \epsilon_c)$.  In this case the distance
between the two roots of Eq. (\ref{QUADEQ}) is proportional to 
$(\epsilon- \epsilon_c)^{1/2}$, and thus the distance between the corresponding
value of $\rho$ is also proportional to $(\epsilon- \epsilon_c)^{1/2}$, which
is proportional to $T_c-T)^{1/2}$.  Examination of Fig. \ref{FIG3}, without
additional analysis, makes it clear that the difference in density between
the two coexisting phases, $y_3(z_0) - y_1(z_0)$ is also proportional to
$(T_c-T)^{1/2}$.

\section{Mean Field Theory}
\label{MFAP}

In this appendix we obtain mean field theory for interacting dimers
from the Hamiltonian using the standard decoupling even though
this Hamiltonian involves operators which obey unusual trace rules.

In view of the trace rules (extended to infinite $q$) we may write
\begin{eqnarray}
e^{- \beta {\cal H}} &=& \prod_{\langle ij \rangle} e^{\hat z s_i s_j + \alpha \beta
t_i t_j} \ ,
\end{eqnarray}
where $\hat z = z e^{-\beta \alpha}$.  Mean field theory is obtained by
ignoring correlated fluctuations and writing
\begin{eqnarray}
- \beta {\cal H} &=& \sum_{\langle ij \rangle} \Biggl[
\hat z \left( s_i \langle s \rangle + s_j \langle s \rangle
- \langle s \rangle^2 \right) \nonumber \\
&& \ + \alpha \beta \left( t_i \langle t \rangle + t_j \langle t \rangle
- \langle t \rangle^2 \right) \Biggr] \ .
\end{eqnarray}
Then
\begin{eqnarray}
\langle s \rangle &=& { {\rm Tr}_i s_i e^{s_i q \hat z \langle s \rangle
+ q \alpha \beta \langle t \rangle t_i} \over
{\rm Tr}_i e^{s_i q \hat z \langle s \rangle
+ q \alpha \beta \langle t \rangle t_i} } \nonumber \\
&=& { e^{q \alpha \beta \langle t \rangle} \over
1 + q \hat z \langle s \rangle e^{q \alpha \beta \langle t \rangle } } 
\label{SAVEQ}
\end{eqnarray}
\begin{eqnarray}
\langle t \rangle &=& { {\rm Tr}_i t_i e^{s_i q \hat z \langle s \rangle
+ q \alpha \beta \langle t \rangle t_i} \over
{\rm Tr}_i e^{s_i q \hat z \langle s \rangle
+ q \alpha \beta \langle t \rangle t_i} } \nonumber \\
&=& { q \hat z \langle s \rangle e^{q \alpha \beta \langle t \rangle} \over
1 + q \hat z \langle s \rangle e^{q \alpha \beta \langle t \rangle } } \ . 
\label{TAVEQ}
\end{eqnarray}
We solve Eq. (\ref{TAVEQ}) for $\langle s \rangle$ to get
\begin{eqnarray}
\langle s \rangle &=& {\langle t \rangle \over q\hat z e^{q \alpha \beta
\langle t \rangle} [1 - \langle t \rangle] } \ ,
\end{eqnarray}
so that Eq. (\ref{SAVEQ}) gives
\begin{eqnarray}
\ln \langle t \rangle = \ln (q \hat z) + 2 q \alpha \beta \langle t \rangle
+2 \ln [1 - \langle t \rangle ] \ .
\end{eqnarray}

To shorten the discussion we assume a second order transition in which case
the above equations {\it and} its two derivatives with respect to $\langle t \rangle$
are zero.  Differentiating twice we get
\begin{eqnarray}
{1 \over \langle t \rangle } = 2q \alpha \beta - {2 \over 1 - \langle t \rangle }
\end{eqnarray}
and
\begin{eqnarray}
- {1 \over \langle t \rangle^2} = - {2 \over (1 - \langle t \rangle)^2 } \ .
\end{eqnarray}
The last equation gives $t_c$, the value of $\langle t \rangle$ at the
critical point, to be $t_c = \sqrt 2 -1$.  Putting this into
the preceding equation gives $q \alpha \beta_c = \sqrt 2 + (3/2)$.

We may compare this with the solution for the Bethe lattice.  For simple models
(such as the Ising model) it is known that the Bethe lattice solution only
coincides with mean field theory for large $q$.  For large $q$ the Bethe lattice
solution for $T_c$ given by Eq. (\ref{ECEQ}) agrees with the present result.

\section{Geometrical Calculation of First Loop Correction for
the Triangular Lattice}

For the triangular lattice, a simple geometrical argument 
permits us to calculate most of the difference between the "exact" 
value of  $\overline N_D /N$  (notation as in Table I) and the 
value given by the Bethe approximation. The argument depends on the 
"tight" structure of the triangular lattice, and we have not been 
able to extend the argument to the square lattice. Furthermore, 
unlike the "loop corrections"  calculated in  Sec. IIC, our 
geometrical calculation is  not the first term in a systematic series 
of corrections.  Nevertheless, the calculation is simple and 
remarkably accurate.  From Eq. (7) we have
\begin{eqnarray}
1/z = \rho_V /\rho  \ ,
\label{C1} \end{eqnarray}
where $\rho_V$ is the probability that adjacent sites A and B are 
both vacant and $\rho$ is the probability that a bond is occupied. 
Clearly
\begin{eqnarray}
P({\rm B \ vacant}) & =&  P({\rm B \ vacant \ and \ A \ vacant})
\nonumber \\ && \ + P( {\rm B \ vacant \ and \ A \ occupied}) \ .
\label{C2} \end{eqnarray}

\vspace{0.2 in}
\begin{figure}
\centerline{\psfig{figure=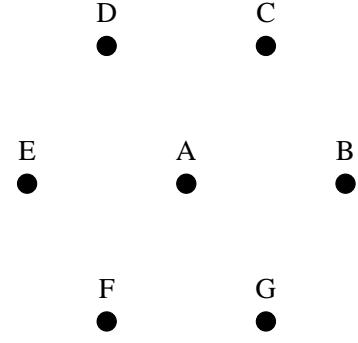}}
\caption{A portion of the triangular lattice.}
\label{MCT.FIG} \end{figure}

\vspace{0.2 in}
Consider the section of the triangular lattice shown in Fig. \ref{MCT.FIG}.
We see that
\begin{eqnarray}
P({\rm B\ vacant}) &=& P({\rm A \ vacant}) =  1 - 6 \rho
\end{eqnarray}
and we set
\begin{eqnarray}
P({\rm B\ vacant\ and\ A\ vacant})=(1- 6 \rho) (p') \ ,
\end{eqnarray}
where
\begin{eqnarray}
p' &=& P({\rm B\ vacant\ | \ A \ vacant})\  .
\end{eqnarray}
Furthermore
\begin{eqnarray}
&& P({\rm B \ vacant\ and\ A\ occupied}) \nonumber \\ &&
= P({\rm B\ vac\ and\ AE\ occ})
+ P({\rm B\ vac\ and\ AD\ occ})
\nonumber \\ && \ \ + P({\rm B\ vac\ and\ AF\ occ})
+ P({\rm B\ vac\ and\ AC\ occ})
\nonumber \\ &&\ \ + P({\rm B\ vac\ and\ AG\ occ})\ .
\label{5TERM} \end{eqnarray}
We now estimate each of the terms on the right-hand side
of this equation.  It seems evident that there is very little
difference between $P({\rm B \ vacant} \ |\ {\rm AE \ occ})$ and
$P({\rm B \ vac |\ A \ vacant})$, since the influence of the
extra dimer would have to propagate over a long tortuous path.
Thus we write $P({\rm B\ vac\ and\ AE\ occ})= (\rho)(p')$.
Similarly, and on slightly weaker ground, we write
\begin{eqnarray}
&& \ P({\rm B\ vacant\ and\ AD\ occupied}) \nonumber \\
&=& P({\rm B\ vacant\ and\ AF\ occupied})=(\rho)(p')\ .
\end{eqnarray}
To estimate $P({\rm B\ vac \ and \ AC \ occ})$ we note that 
if AC is occupied then AB and CB must be unoccupied. As far as
state of site B is concerned, it makes little difference whether
we specify that bond AC is occupied or that bonds AB and CB are
unoccupied, because the effect of the presence of the dimer AC
on the state of site B is very indirect.  Thus we make the approximation
that
\begin{eqnarray}
&& \ P({\rm B\ vacant\ |\ AC\ occupied}) \nonumber \\
&=& P({\rm B\ vacant\ |\ AB \ and \ BC \ unoccupied}) \ .
\end{eqnarray}
However, the
conditional probability that B is vacant, given that AB and CB are 
unoccupied, is $(1-6\rho)/(1-2\rho)$, so that approximately
\begin{eqnarray}
P({\rm B\ vac \ |\ AC\ occ}) &=&
(1-6\rho)/(1-2\rho) \  ,
\end{eqnarray}
and thereby that
\begin{eqnarray}
P({\rm B\ vac \ and \ AC\ occ}) &=& \rho (1-6\rho)/(1-2\rho) \  .
\end{eqnarray}
Now we use our evaluation of each term on the right-hand side of
Eq. (\ref{5TERM}) to write Eq. (\ref{C2}) as
\begin{eqnarray}
(1-6\rho) &=& (1-6\rho)(p')+(3\rho)(p') \nonumber \\ && \ \
+(2\rho)(1-6\rho)/1-2\rho)\ . 
\end{eqnarray}

Solving for  $p'$ we find that
\begin{eqnarray}
p' &=& (1-6\rho)(1-4\rho)/(1-2\rho)(1-3\rho)
\end{eqnarray}
and Eq. (\ref{C1}) yields
\begin{eqnarray}
1/z &=& (1-6\rho)(p')/(\rho) \nonumber \\ &=&(1-6\rho)^2 
(1-4\rho)/(\rho)(1-2\rho)(1-3\rho) \ ,
\end{eqnarray}

The computer readily calculates $\rho$ for each value of $z$ in Table 
II. The corresponding value of  $\overline N_D/ N$, which we call
$\pi_G$, is $3\rho$ and is exhibited in Table II.

Thus, a simple geometrical argument yields a good estimate of the 
"first loop correction" for the triangular lattice. However, we have 
been unable to make a corresponding argument for the square lattice, 
nor can we extend this  argument in an orderly way to make higher 
order corrections.
\end{appendix}


\begin{thebibliography} {}
\bibitem{FN4}
Either or both subsystems of bonds S$_1$  and S$_2$ may include 
bonds which intersect the boundary of the tree.
\bibitem{FN2}
Implicitly we treat the independent subsystems
T$_{\rm A}$ and T$_{\rm B}$ via
a grand canonical formulation in which each dimer carries an
activity $\zeta$, which has to be adjusted to give the
equilibrium value of the total number of dimers in the combined
system.  In equilibrium, the correct value of $\zeta$ is indeed
$z \exp[\beta \alpha (2q-1)]$.  However, the final result
for $p_{\rm con}$ in Eq. (\ref{POFYEQ}), below, does not involve $\zeta$.
Therefore we can use Eqs. (\ref{INTEQ3}) and (\ref{POFYEQ}) for
values of $N_D$ which are not equal to $\overline N_D$,
as we do in Appendix \ref{VDWAP}.
\bibitem{1939}
An equation equivalent to Eq. (\ref{SELFC}) was obtained in
Ref. \onlinecite{B1}, but was not related to the Bethe lattice.
\end{thebibliography}
\end{document}